\documentclass[12pt,preprint]{aastex} 

\newcommand{\simgt}{\hbox{\rlap{\raise 0.425ex\hbox{$>$}}\lower 0.65ex\hbox{$\sim$}}}
\newcommand{\simlt}{\hbox{\rlap{\raise 0.425ex\hbox{$<$}}\lower 0.65ex\hbox{$\sim$}}}

\newcommand{\s}{$\phantom{p}$}

\newcommand{\etal}{{\it et al.\ }} 
\newcommand{\eg}{{\it e.g.\ }}

\shorttitle{Dark matter halos}   
\shortauthors{Williams \etal} 

\begin{document}   

\title{Investigating the Origins of Dark Matter Halo Density Profiles}

\author{Liliya L.R. Williams}
\affil{Astronomy Department, University of Minnesota, Minneapolis, MN 55455} 
\email{llrw@astro.umn.edu} 

\author{{Arif Babul}\altaffilmark{1}}
\affil{Department of Physics and Astronomy, University of Victoria, BC, Canada} 
\email{babul@uvic.ca} 
\altaffiltext{1}{CITA Senior Fellow}

\author{{Julianne J. Dalcanton}\altaffilmark{2}} 
\affil{Astronomy Department, University of Washington, Box 351580, Seattle, WA 98195} 
\email{jd@astro.washington.edu} 
\altaffiltext{2}{Alfred P. Sloan Foundation Fellow} 

\begin{abstract} 
Although high-resolution N-body simulations make robust empirical 
predictions for the density distribution within cold dark matter 
halos, these studies have yielded little physical insight into the origins 
of the distribution. We therefore attempt to investigate the problem using 
analytic and semi-analytic approaches. Simple analytic considerations suggest 
that the inner slope of the central cusps in dark matter halos
cannot be steeper than $\alpha =2$ (where 
$\rho\propto r^{-\alpha}$), with $\alpha=1.5$--$1.7$ being a more realistic 
upper limit. Moreover, our analysis suggests that any number of effects, 
whether real (eg. angular momentum imparted by  tidal torques
and secondary perturbations) or artificial (eg. two-body interactions,  
the accuracy of the numerical integrator, round-off errors), will result in 
shallower slopes. We also find that the halos should exhibit a well-defined 
relationship between $r_{peri}/r_{apo}$ and
$j_{\theta}/j_r$.  We derive this relationship analytically and speculate 
that it may be ``universal''.  Using a semi-analytic scheme based 
on Ryden \& Gunn (1987), we further explore the 
relationship between the specific angular momentum distribution in a halo 
and its density profile. For present purposes, we restrict ourselves to halos 
that form primarily via nearly-smooth accretion of matter, and only consider 
the specific angular momentum generated by secondary perturbations
associated with the cold dark matter spectrum of density fluctuations.   
Compared to those formed in N-body simulations, our ``semi-analytic'' halos 
are more extended, 
have flatter rotation curves and have higher specific angular momentum, even 
though we have not yet taken into account the effects of tidal torques.     
Whether the density profile of numerical halos is indeed the result
of loss in angular momentum outside the central region, and whether 
this loss is a feature of hierarchical merging and major mergers in 
particular, is under investigation.
\end{abstract} 


\section{Introduction} 

Over the past two decades, observational and theoretical progress have  
given rise to an increasingly detailed picture of how structure in the
Universe forms and evolves.  In the prevailing paradigm for the
formation of structure in the Universe, galaxies and clusters of
galaxies are contained within extended cold dark matter (CDM) halos,
formed hierarchically through gravitationally induced mergers of
smaller-scale structure.  The evolution of individual dark
matter halos and the formation of galactic structure inside
the halos is strongly dependent on a daunting array of physical
processes that span a wide range of scales.  It is not surprising,
therefore, that even the very first steps in the assembly of
galaxies -- the details of how dark halos form and virialize, the
shape and profile of their equilibrium structure, and the extent to
which that structure is shaped by the surrounding large-scale
structure and the halo's merger history -- are not well understood.

At the simplest level, dark matter halos form when, in the early
Universe, the matter within (and surrounding) an overdense region
suffers gravitational retardation, decouples from the Hubble flow,
collapses, and in due course, virializes.  The basic theoretical
framework underlying the
above sequence was outlined by Gunn \& Gott (1972), Gott (1975) and
Gunn (1977), and subsequently elaborated upon by Fillmore \& Goldreich
(1984) and Hoffman \& Shaham (1985).  In order to make the problem
analytically 
tractable, several simplifications had to be invoked, including
spherical symmetry, purely radial particle orbits, and power-law
initial density profiles.  These studies predicted power-law halo
density profiles, $\rho(r)\propto r^{-\alpha}$, with index $\alpha$
ranging from $2$ to $2.25$. These initial results were encouraging in that 
the density profile, at the very least, could explain the nearly flat 
rotation curves of massive spiral galaxies.

In a bid to relax the restrictive assumptions underlying the analytic
investigations and to incorporate the full range of non-linear
gravitational effects, Quinn, Salmon \& Zurek (1986) pioneered the use
of N-body simulations to study halo formation.
Subsequent higher resolution simulations by Frenk \etal (1988),
Dubinski \& Carlberg (1991),
Navarro, Frenk \& White (1996, 1997, hereafter NFW), Moore \etal
(1998) Jing \& Suto (2000), Klypin \etal (2001), Bullock \etal (2001),
and Power \etal (2003)
found that although the spherically-averaged density profiles of the
N-body dark matter halos are similar regardless of the mass of the
halo or the cosmological model, they are significantly different from
the single power laws predicted by the theoretical studies.  The
N-body profiles are characterized by an $r^{-3}$ decline at large
radii and a cuspy profile of the form $\rho(r)\propto r^{-\alpha}$
where $\alpha < 2$ near the center. The actual value of the inner density 
slope $\alpha$ is a matter of some controversy, with NFW and 
Power \etal (2002) suggesting $\alpha=1$, but with Moore \etal
(1998), Ghigna \etal (2000) and Fukushige \& Makino (2001) arguing for
$\alpha=1.5$. 
Klypin \etal (2001) argue that NFW and Moore \etal
profiles are compatible with each other, within their range of applicability:
the profile can only be trusted beyond radii which are 2-4 times larger than 
the formal resolution of the simulation. 

In spite of the controversy about the exact slope of the cusp at the
center of the simulated halos the ``universal''
N-body density profiles have been widely adopted for theoretical work.
The analytic form for the profile is simple and easy to work with, and
does have some observational support on the scale of clusters (\eg
Carlberg \etal 1997).  However, there are two important 
unresolved issues that require attention.

While numerical simulations universally produce the
aforementioned cuspy density profiles, nature apparently does not.
The predicted profiles appear to disagree with observed dark matter
profiles, in both overall density and in shape, particularly at
smaller mass scales.
Some of the strongest evidence of the failure of the universal
N-body profiles comes from observations of the dynamics of spiral
galaxies.  There is mounting evidence from rotation curve studies that 
unlike galaxies in simulations, late-type galaxies may be embedded within
halos that have shallow dark matter cores (Salucci 2001; Salucci
\& Burkert 2000; Hernandez \etal 2001).
Within low surface brightness (LSB) galaxies, the difference between
the observed rotation curves and those predicted by N-body halo
profiles becomes even more severe 
(\eg Swaters \etal 2000; Dalcanton \& Bernstein 2000;
van den Bosch \& Swaters 2001; de Blok \etal 2001; de Blok \& Bosma 2002; 
Marschesini \etal 2002; Weldrake \etal 2002; Swaters \etal 2002);
the cores of these galaxies are much less dense
than the simulations indicate.  These LSB galaxies are thought to be 
ideal for the comparison with theory, as their dynamics are dominated 
by dark matter with little contribution from baryons (Bothun \etal 1997), 
and thus the discrepancy with simulations is particularly troublesome.

Quite apart from the apparent discrepancies between numerical results
and observations, there is yet another problem, on which very little light 
has been shed in spite of the tremendous progress achieved by numerical 
work: What is the physics underlying the universal density profiles 
predicted by simulations? There have been several attempts in the 
literature to explain the shapes of virialized halos. Intuitively, one 
could argue that the major difference between earlier analytical work 
and numerical simulations is that halo formation in N-body simulations 
proceeds through repeated mergers. In fact, Syer \& White (1998) argue 
that universal N-body profile is a natural outcome of hierarchical 
merging.  Specifically, they claim that the halo structure in the inner 
regions is determined by a nested sequence of undigested cores of merging 
substructure accumulating in the centers of the halos.
However, Huss \etal (1999) find that simulations of isolated halos collapsing
more or less spherically also result in universal profiles,
thus suggesting that hierarchical merging is not crucial to the outcome.  
Instead, they suggest that the profile is a consequence of a near universal 
angular momentum distribution of the halos. Unfortunately, it is unclear what 
circumstances lead the halos to this universal angular momentum distribution. 
Thus the issue remains unresolved. 

Our current state of knowledge about halos from theoretical work
can be summed up as follows: universal halos are a very robust prediction
of numerical N-body simulations, but we do not understand how their properties 
arise. This current lack of understanding of numerical halos, and hence the 
lack of understanding of observed galactic halos, provides the main impetus 
for our present and subsequent work.

In this paper we describe our dynamical method for generating virialized
galaxy-scale halos, and address some basic questions regarding halo formation. 
Since the very large amount of work carried out by many researchers to date 
using N-body simulations has met with limited success in elucidating the
physics of halo formation, it makes sense to consider a different approach: 
we use an analytical technique, pioneered by Ryden \& Gunn (1987, RG87), 
which we describe in detail in Section~\ref{RG87} and Appendix~\ref{adnauseum}.

The most important feature of the original RG87 method is that the dynamical 
evolution is carried out while conserving angular and radial momenta of 
individual halo shells. 
The initial shape of the proto-halos is derived from the fluctuation spectrum 
at high redshifts, and halos are endowed with secondary perturbations, which
impart random motions to halo particles. The statistical properties of the 
secondary perturbations are derived from 
the same fluctuation power spectrum, and therefore their effects on 
particles' random velocities are treated self-consistently. 

In this paper we extend the RG87 method in order to more faithfully reproduce 
the real halos, and to allow experimentation, using a range of possible 
cosmological conditions.
Compared to the original RG87 formalism, our improved formalism allows
the particles at a given radial location in the halo to acquire a range of 
velocities, not just a single rms value. The velocity distribution 
can be specified independently, allowing us to either constrain
the distribution to be consistent with the underlying statistics of a specific
dark matter model or to experiment with different distributions in order to
understand the impact of non-radial and radial motions on the final structure 
and dynamical properties of halos.

The current version of our method treats collapse and virialization of halos 
that are spherically symmetric, that have suffered no major mergers or 
tidal torquing, and that have experienced only quiescent 
accretion of somewhat lumpy material, i.e. minor mergers. In this work
we assume that all parts of the main halo experience statistically the same
secondary perturbations, in other words, the secondary perturbation fluctuation 
field in unconstrained. 
Because our halos cannot undergo major mergers, they do not form 
in a fully hierarchical setting. However, the quiescent accretion scenario
is worth investigating: judging by the commonness of extended
thin spiral disks in the Universe, it is not far fetched to suppose that
such formation histories are realistic, at least for some galaxies.

Of course, halo formation in a hierarchical scenario is a highly non-linear 
process. An analytical method, no matter how sophisticated, will never be 
able to capture the full extent of complexity of a non-linear process.
And therein lies the power of the analytical approach: numerical simulations 
yield little physical insight beyond empirical findings precisely because 
they are so rich in dynamical processes, which are hard to disentangle and 
interpret in terms of underlying physics. 
We aim to complement the detailed picture of halo formation as revealed by
simulations with the understanding of its broad-brush features.
Put differently, we are seeking an impressionist's view of halo formation, 
not that of a Dutch master. 

The paper is organized as follows. In Section~\ref{synopsis} we give a broad
overview of the formation of dark matter halos, and quote some major analytical 
results from previous work. Section~\ref{RG87} briefly describes the original 
Ryden \& Gunn (1987) method, which we use a starting point for our work; a
fuller description can be found in the Appendix.
In Section~\ref{whatwedid} 
we generate virialized halos for a range of galaxy-sized halo masses. We
obtain rotation curves and specific angular momentum distribution of our halos,
and discuss how these relate to their N-body and observed counterparts.  
We also explore the role that the initial conditions play in 
determining the final density profile and halo dynamics. To that end, 
we consider fluctuation power spectra that are tilted compared to the 
Harrison-Zel'dovich spectrum, or are truncated at short wavelengths. 
In Section~\ref{understanding} we use the results of these experiments, 
combined with analytical reasoning to provide answers to some general 
questions about halos: why central cusps have slopes around 1-1.6, why the
slope of log-log density profiles steepens monotonically from the center
outward, and why virialized halos formed as a result of quiescent accretion
retain information about the shape of the primordial power spectrum.

\section{Analytical treatments of dark matter halos}\label{synopsis}

To put our analytical method in context we first present a broad picture of
halo formation, emphasizing some of the relevant analytical work carried out 
to date. 

In the last three to four decades a considerable amount of analytical work 
has been carried out to understand the formation of bound structures through 
gravitational instability. A number of approaches have been tried, but because 
of natural limitations it is difficult for any one analytical formalism 
to address all aspects of galaxy formation. Most papers in the literature 
concentrate on a specific set of initial conditions, or treat only a limited 
range in radii of halos. Consequently, comparison between different results 
is not always straightforward. Here we attempt to put together a 
coherent picture of the past results. 

Consider an isolated halo, which has had no major mergers in its history, and 
divide it roughly into four regions starting from the outside: In Region (1),
furthest from the center of the initial density peak, dark matter particles are 
beginning to feel the gravitational tug of the central peak and are just 
starting to fall behind the Hubble flow. Further in, in Region (2), 
the particles are starting to decouple from the Hubble flow and are about 
to begin collapse. In Region (3) the central density peak dominates the 
motion of particles; this is the region of infall and shell crossing. 
Finally, in Region (4), the central part of the density peak, virialization 
is taking place, or has already been reached.
Note that the evolution time sequence of the halo maps roughly on to
the distance away from the halo center, but the demarcation of various regions, 
both in time and radial distance, is only approximate. As time progresses the
boundaries of all the regions in an isolated halo move out
from the center, as viewed in initial comoving coordinates, i.e. the halo
engulfs an ever increasing fraction of comoving volume. In an Einstein-de Sitter
Universe this process can proceed indefinitely, while in a less dense universe
the accretion of material is limited.

Region (1) consists of the still expanding material. Assuming spherical
symmetry, Newton's first theorem says that the evolution in
this region is completely determined by the matter interior to the shell.
The evolution obeys a matter-dominated Friedmann equation, with the 
density $\Omega_{matter}$ equal to the average density inside the shell. 
While the shells are still expanding the well known parametric solution applies 
(Gunn \& Gott 1972, Peebles 1980), given in equations \ref{rtheta} and \ref{ttheta} 
of the Appendix.
The applicability of the solution ends when a given shell reaches its
turn-around radius at the time when the average interior density contrast 
is $\sim 4-5$. The corresponding extrapolated linear density contrast is 
$\sim 1$. 

Region (2), which contains decoupled and infalling material, starts roughly 
where the linear extrapolated density contrast is of order 1. 
Region (2) consists of material that has gone past turnaround
and is infalling for the first time, but is sufficiently far from virial
radius that there is no shell crossing occurring, only infall. The typical
densities in this region range from 5-100 $\times\rho_{crit}$.

In Region (3), material is falling inwards,
encountering inner shells that are expanding, or recollapsing; 
this is the shell-crossing region. Some of the shells are falling in 
for the first time; others may have undergone one or more oscillations but 
have not relaxed. It is well characterized by the scaling solutions of 
Gott (1975), and Gunn (1977), for which the slope of the density profile 
is a function of the ambient $\Omega_{matter}$, with steeper slopes expected 
for smaller $\Omega_{matter}$. In these models,
critical density Universe produces halos with slopes of $2.25$. 
Fillmore \& Goldreich (1984) adopt a power law initial density profile 
and use a better dynamical approximation for treating the shells
past the turnaround, leading to final profile slopes from 
$\rho\propto r^{-2.25}$ to $\rho\propto r^{-2}$. The typical 
densities in this region are about $100-500\times \rho_{crit}$, and since
$t_{orbital}\sim H_0^{-1} \sqrt 8\;\pi [\rho/\rho_{crit}]^{-1/2}$
(or, for Einstein-de Sitter, $t_{orbital}/t_0 \sim 13.5 [\rho/\rho_{crit}]^{-1/2}$)
a typical particle in this region has completed at least one passage 
through the halo. Results of RG87 show that the density slopes
at density contrasts of $100$ are about $2$.

In Region (4), virialization is well underway or is complete. 
Many analytical solutions in regions (1)-(3) ignore non-radial motions of 
particles; all orbits are assumed to be purely radial. This is probably a 
tolerable approximation for these regions. However, the approximation most 
definitely breaks down in the virialized central region of the halo, Region (4). 
Though  Gott (1975), Gunn (1977), and Fillmore \& Goldreich (1984) argue
that their solutions apply to the virialized regions, this is not
the case. Even in very smooth initial halos, particles
will have some non-zero tangential velocities. These will become increasingly 
important as particles start falling towards the center of the halo.
For a particle falling towards a centrally condensed mass, 
the radial velocity $v_{rad}\propto r^{b}$, where $b\ge -1/2$,
with the exact value depending on the density distribution. 
The particle's angular momentum $r\; v_{tan}$, stays constant. 
As a result, $v_{tan}/v_{rad}\propto r^{c}$, where $c\le -1/2$. 
Thus, tangential velocities always increase as particles fall inward.
The calculations carried out in RG87 demonstrate that non-radial motions are 
very important in determining the outcome in this region. Non-radial motions
give particles angular momentum, which prevents them from penetrating to the
very center of the halo. This picture is supported by the recent work of 
Avila-Reese, Firmani \& Hernandez (1998), Subramanian, Cen \& Ostriker (1999) 
and Hozumi, Burkert \& Fujiwara (2000), among others. For example, 
Subramanian \etal apply 
the collisionless Boltzmann equation to self-similar halo collapse to show 
that tangential velocities are necessary to attain central slopes
shallower than $2$. Hozumi \etal start with a power law density profile and
an initial $v_{tan}/v_{rad}$ ratio and integrate the collisionless Boltzmann 
equation to arrive at the final phase-space density distribution of the halo. 
Hozumi \etal and others stress the importance of the 
initial distribution of radial vs. tangential velocities in the halo in
determining the final profile of the central region of the halo.

The rates at which matter in regions (1), (2), (3) and (4) makes its 
way towards the center of the halo are different, reflecting the different
types of dynamical processes going on. Therefore one should not expect 
a single density profile power law index to hold throughout the halo.
In fact, all analyses that do not explicitly assume that power law profiles 
hold over a wide range of radii, find that profile slope does change with 
radius. 

The dynamical process described above, that of accretion of mass from an
expanding background and growth of structures from inside out, is called the 
secondary infall model (SIM). The assumptions used in the analytical 
implementation of SIMs were highly idealized in the 1970's, but have become 
more realistic in the recent years. For example, 
early models started with a $\delta$-function or power law seed structure, and
assumed purely radial particle motions and self-similar collapse, while more 
recent models relax all these assumptions.  Table~\ref{table1}
summarizes some of the key SIM papers, and documents how the model
assumptions have changed over time. An underlined item in 
Table~\ref{table1} means that the corresponding paper was the first published
use of that particular assumption (to the best of our knowledge).

SIM models were first explored
in detail by Gunn \& Gott (1972, GG72) who studied the evolution
of a spherically symmetric halo up to the time of the maximum expansion. 
Gunn (1977, G77) followed the evolution of halos past the maximum expansion
and into the shell-crossing regime. This paper
introduced the use of adiabatic invariants: for self-similar radial collapse 
radial momentum is equivalent to $r\cdot m(r)$, where $m(r)$ is the mass 
enclosed within $r$. Fillmore and Goldreich (1974, FG84) further explored 
the dynamics of self-similar halo collapse, using a power law as the initial 
density profile. 
Zaroubi \& Hoffman (1993, ZH93) refined the FG84 collapse calculations, 
by improving the estimation of the collapse factor of shells, i.e. the
ratio of the shell's radius of maximum expansion and its asymptotic radius.
Hoffman and Shaham (1985) used the
two-point correlation function as a more realistic approximation to
the initial density profile of halos.
However, the radial shape of the typical {\it overdense} regions, i.e. peaks
in the density field is steeper than the correlation function. 
Ryden \& Gunn (1987) were the first to adopt peaks in the density field as the
density profile of the proto-halos. RG87 were also the first 
to relax the assumption of purely radial self-similar collapse by including
non-radial motions arising from secondary perturbations in the halo.
Inclusion of non-radial motions required an additional adiabatic invariant;
RG87's model conserves angular and radial momenta of each shell 
throughout the collapse.
Avila-Reese \etal (1998) differs from most other works because
their halos are assembled through a series of discrete merger episodes,
described by mass aggregation histories (MAH), which are generated from 
the Gaussian fluctuation field. Their dark matter particles are endowed with 
``thermal motions'' resulting in non-radial velocities. Nusser (2000)
tried two different analytical schemes for non-radial velocities in his halos,
and concluded that angular momentum is most effective when added to a
particle at the time of maximum expansion.
Most of the SIM treatments assume that any given halo is completely isolated
and so continues to accrete mass indefinitely, however Lokas (2000) argue that 
mass can be accreted only from a finite sphere, whose radius is 1/2 of the 
typical peak-to-peak separation. This results in the steepening of the outer 
density profile of the final halo.
The last column in Table~\ref{table1} states whether the final virialized
density profile is a power law or not. All non-power law profiles have
a `concave' shape in the log-log plane.

Several of the papers listed in Table~\ref{table1}, as well as other works
(e.g. Del Popolo \etal 2002, Hiotelis 2002, Kull 1999, Lokas \& Hoffman 2000) 
compared their results to those of N-body simulations. In most cases 
they found their final profiles to be similar to NFW empirical fit, in particular,
analytically generated halos possess central density cusps. In contrast, our density
profiles, which we describe in Section~\ref{whatwedid}, often have small flat 
density cores. This difference probably arises from the different methods
used to generate the angular momenta of particles. We assume that angular 
momenta arise from the random secondary perturbations, and derive their 
magnitude in a self-consistent fashion from the same fluctuation power spectrum 
that generates the primary peak. All other analytical treatments use simple 
prescriptions to obtain angular momenta of particles as a function of radius, 
for example, $L^2 \propto M(<r)\, r$.

\section{A brief summary of the Ryden \& Gunn 1987 method}\label{RG87}

Our halo formation formalism builds upon the powerful analytical method 
introduced by Ryden \& Gunn (1987), which allows one to track the evolution 
of dark matter halos in all four regions discussed in Section~\ref{synopsis}, 
while accounting for the effects of non-radial motions and shell crossing.
In this Section we summarize the key elements of the original RG87 method;
our refinements and extensions are discussed in Section~\ref{whatwedid}. 
All the technical details are relegated to the Appendix.

In most popular cosmological scenarios the density field
soon after recombination can be represented by a Gaussian random field.
The amplitude of density fluctuations at that early time is very small, 
$\sim 10^{-5}-10^{-4}$.
If the initial spectrum is close to Harrison-Zel'dovich then the early
evolution of the density field is very well described by linear theory,
where, in the Einstein-de Sitter case the amplitude of perturbations grows
as $\delta\rho/\rho\propto a(t)\propto t^{2/3}$, i.e. our Region (1).
High density contrast peaks in the 
field will eventually achieve overdensities of order 1 and enter a non-linear
stage of evolution. These peaks will then collapse to form bound structures. 
We start with one of these peaks, and, for simplicity take it to be 
spherically symmetric. The peak is divided into a very small central core
and many spherically symmetric concentric mass shells, each labeled by its initial
comoving distance from the center, $x$. Each shell's evolution is divided into
two stages. Initially a shell expands with the Hubble flow, but with a slight 
deceleration arising from the central mass concentration. Eventually the shell's 
outward radial velocity decreases to zero, after which the shell collapses, 
by some distance, back towards the center of the halo. The dividing moment is 
called the turn-around, and at any given time corresponds to the line dividing 
Regions (1) and (2). In a halo with a declining density profile the 
turn-around happens at progressively later cosmic times for shells at greater 
distances from the center.

In reality, the initial density peak will not be smooth, but will instead be 
sprinkled with many smaller scale positive and negative perturbations that arise 
from the same Gaussian random field
that gave rise to the main peak. These secondary perturbations will perturb the 
motion of the dark matter particles from their otherwise purely radial orbits.

In the early part of the evolution of the halo most of the shells are still expanding.
During this time secondary perturbations grow, and so does the acceleration, the 
velocity, and the displacement contributed by these perturbations to the 
particles in the shell. Because the secondary peaks are randomly distributed 
within the halo, they displace the dark matter  particles in random directions 
from their original positions. This can be visualized as a shell having an 
internal velocity dispersion, resulting in a `puffy' shell. If we concentrate 
on a single particle, its orbit, viewed from the rest frame of the parent shell, 
will oscillate between an inner and an outer radius of that shell (i.e. peri- and 
apocenter). In general, this orbit will not be closed and will resemble a rosette. A time 
averaged picture of one particle's orbit in this scenario will be identical to a 
single-moment snapshot of all the particles in a shell\footnote{Note that in the
rest of the paper we use the words `shells' and `particles' interchangeably.}.

In a real situation, the puffiness of any given shell will gradually increase as the 
shell expands away from the halo's center, and the influence of the secondary 
peaks grows, but for simplicity, our calculations assume that each shell 
stays dynamically cold and spatially thin during the expansion, and then puffs up 
instantly as it reaches turnaround. In other words, as long as a given shell is 
expanding its dark matter particles' positions 
and velocities are not corrected for the effects of secondary perturbations; 
it is assumed that during this relatively orderly evolutionary stage the effects
of random perturbations are small compared to what they will be later when
the halo starts to collapse.  So, the contributions are evaluated analytically
but are not imparted to the shell until it reaches turnaround.
Furthermore, only the velocities of dark matter particles are corrected; 
the positions are left unchanged, so in our halos the puffiness of the shells 
refers to their dynamical state, not to the spatial distribution of their dark 
matter particles.

At turnaround, every particles' kinetic energy is augmented by the additional 
velocity component, $\Delta {\bf v_{\rm rms}}(x)$, the rms value of the accumulated 
effect of secondary perturbations, which is a function of radial position in the 
halo. The three spatial directions are statistically equivalent as far as random 
velocities are concerned, so $1/3$ of the total amplitude of 
$|\Delta {\bf v_{\rm rms}}(x)|^2$ goes to each of the three orthogonal directions. 
Instead of having $\Delta {\bf v}$ be defined uniquely by $x$, and dividing the 
amplitude equally among the three orthogonal directions, as was done in the 
original RG87 method, our formalism implements a distribution of velocity values 
at a given $x$ (using the same $\Delta {\bf v_{\rm rms}}(x)$ as before), 
and picks the orientation of the velocity vector randomly, thereby
better approximating the dark matter velocity structure of real halos.

As every additional shell reaches its turnaround radius and falls in, it
overlaps a number of shells that have already fallen, all of which have
their own peri- and apocenters. This is shell-crossing.
The gravitational potential of the halo changes with every
additional shell falling in, and has to be recalculated at every step.
This means that the shells' orbits that overlap the newly fallen shell also 
have to be recalculated, keeping their angular and radial momenta fixed
throughout the collapse. This process 
of shell readjustment goes on until the present cosmic time is reached.

\section{Halo density profiles}
\label{whatwedid}

In this Section we describe the formation of dark matter halos via our method. 
We start with the adopted structure of the proto-halos, and in Section~\ref{montecarlo} 
describe how our formalism has been extended from RG87.

\subsection{The initial set-up}
\label{set-up}

In the present paper we retain many features of the RG87 method, including the 
Einstein-de Sitter model. We assume  an Einstein-de Sitter background cosmology
instead of using a more popular flat, low density universe because we are most 
interested in the dynamics and spatial distribution of matter in regions 
where the density is well in excess of critical, by 1-5 orders of magnitude.
In such dense regions the global value of $\Omega_{matter}$ should not matter much.

We use the standard CDM power spectrum, i.e. the scale-invariant Harrison-Zel'dovich
$P(k)\propto k$ form. The transfer function that describes the transition of 
the $P(k)$ through the matter-radiation equality epoch was taken from
Bardeen \etal (1996). Instead of RG87's $H_0=75$ km~s$^{-1}$Mpc$^{-1}$ we use 
$H_0=50$ km~s$^{-1}$Mpc$^{-1}$ which corresponds to standard CDM. The spectrum
is normalized such that rms fluctuations in Gaussian smoothed 
spheres\footnote{Note that a Gaussian window of radius $5 h^{-1}$Mpc contains 
the same volume as a top-hat window of radius $8 h^{-1}$Mpc.} of 
radius $5 h^{-1}$Mpc are 0.5, in agreement with several recent determinations
that estimate $\Omega^{\,0.5...0.6}\sigma_8=0.5$ (Eke \etal 1996; Mellier \etal 2001).

The smallest scales considered here are set by the filtering scale $l_0$,
such that 
\begin{equation}
P_s(k)=P(k) e^{-k^2 l_0^2}.
\end{equation}
For consistency, the size of the unresolved core of each dark matter halo
is set to $M_{core}\sim (2\pi)^{3/2}\rho_{crit} l_0^3$. We use 
$M_{core}=10^6 M_\odot$, corresponding $l_0=10$ kpc.
The shape of the core is assumed to be singular isothermal with an outer 
sharp cutoff at $r_{core}\sim 5$ kpc. The exact shape of the core does not 
matter since it is very small, comprising $\sim 10^{-4}-10^{-5}$ of the 
total mass of a typical halo considered here. 
(Section~\ref{four} contains further discussion of the halo's core.)
Note that because the spectrum is filtered below $l_0$, smaller scale 
secondary perturbations are not included in the calculations.
The mathematical details of the halo formation formalism are described 
in the Appendix.

\subsection{Generating final virialized halos}
\label{montecarlo}

We now describe our procedure for generating virialized dark matter halos.
In a real halo, particles of a given shell will have a range of random 
velocities arising from the secondary perturbations. The RG87 formalism allows 
us to calculate the rms value, $|\Delta {\bf v_{\rm rms}}(x)|$
of these random velocities as a function of radius in the halo. RG87 assume 
that every particle in a given shell has the same $\Delta {\bf v}$ value, 
equal to the rms value (Section~\ref{RG87}). We improve upon their method by 
allowing shells to have a {\it distribution} of velocities. The halo is broken 
up into a very large number of shells, and 
for each shell we pick a random value for $|\Delta {\bf v}|$ from a
Maxwell-Boltzmann distribution 
of the appropriate rms half-width. Furthermore, instead of dividing 
$|\Delta {\bf v}|^2$ equally among three spatial directions, as in RG87, we pick 
a random orientation for the velocity vector $\Delta {\bf v}$ for each shell. 
The halo is then evolved in time using the procedure described in Section~\ref{four}. 

Note that in our halos two adjacent shells can have very different 
velocities and orientations of the $\Delta {\bf v}$ vector. This is analogous 
to a real halo, where two nearby dark matter particles 
can have different random velocities, because they have been affected by a
spatially varying field of secondary perturbations. To approximate a real halo, 
our halos are modeled with a very large number of shells. If the number of shells
is not sufficient, then the resulting halo will have an uneven density distribution,
with discontinuous jumps in density: some shells' pericenters will be close to 
halo's center and so will have their mass distributed over a wide range in radii, 
while other shells will be restricted by large angular momentum and will have
their mass piled up in a narrow range of radii. So, two halos with the same input
$\Delta {\bf v_{\rm rms}}(x)$ parameters can have somewhat different virialized 
profiles, especially in the inner halo regions. In other words, there will be
a scatter in final halo properties. To obtain the average, or typical halo,
two approaches can be used:
(1) Generate $n$ halos each having $m$ shells, and then obtain
the `average' halo, which will then have a smooth density profile. 
(2) Generate one halo with  $n\times m$ shells. The single final halo 
has a smooth density profile, because the very large number of shells
smears out irregularities. Both of these approaches give the same results in 
terms of final halo properties, and
serve as a test of the robustness of the method in view
of the randomness of shells' $\Delta {\bf v}$ magnitude and orientation. All profiles 
in the figures presented in this paper were generated using either (1) or (2).

\subsection{The standard reference case}
\label{std}

\begin{figure} 
\plotone{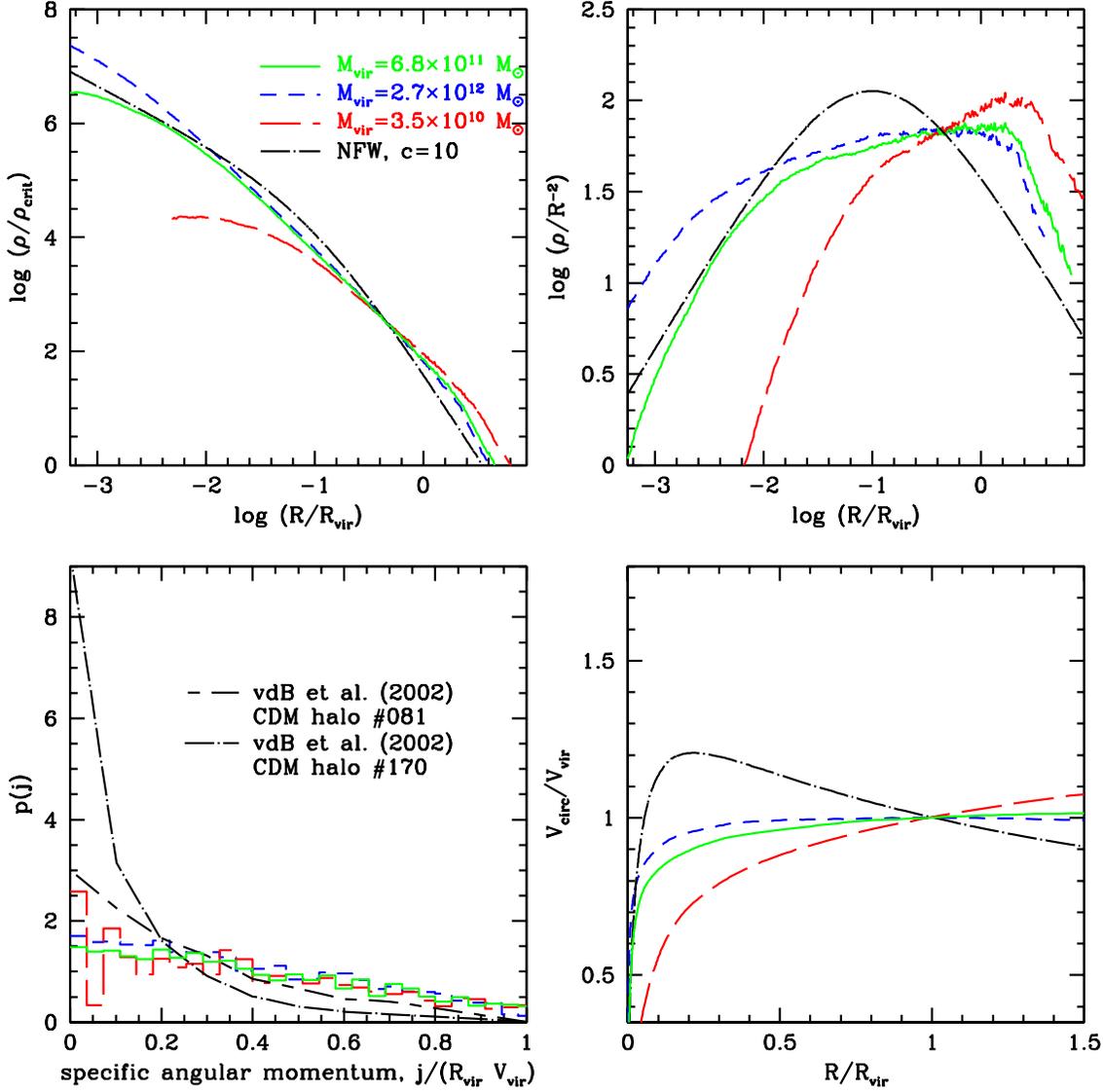} 
\vskip-0.8truecm 
\caption{Dark matter halos generated using ``standard'' initial conditions,
as described in Section~\ref{std}. Halos of three different galactic-type
masses are shown; the intermediate mass halo (solid line) is the `reference'
halo. {\it Upper left:} log-log density profiles. The lowest mass halo 
is not resolved at $\simlt$ 1\% of $R_{vir}$. NFW $c=10$ halo is shown for 
comparison (dot-dash line). {\it Upper right:} density profiles
with the slope of $\alpha=2$ divided out, so $\alpha=2$ slopes are flat 
in this plot. {\it Lower right:} circular rotation curves. {\it Lower left:} 
the distribution of specific angular momenta (SAM) in our halos, 
as histograms. For comparison, we include two extreme SAM distributions 
taken from N-body simulations of van den Bosch \etal (2002).}
\label{small3_std} 
\end{figure}

In this  paper we consider a range of initial parameters and their effect on 
the final virialized halos, focusing on different slopes and filtering 
scales for the primordial matter power spectrum.  
As a baseline, we generate a halo using `standard' assumptions, which 
will be used as a `reference' halo for later results. The reference halo starts 
out as a 3$\,\sigma$ peak in the density fluctuation field of standard CDM spectrum,
which has been filtered with a fixed $l_0$ smoothing scale 
($l_0=10$ kpc, see Section~\ref{set-up})
The virial radius of our reference halo at the present epoch is $R_{vir}=0.23$~Mpc, 
and the mass contained within that virial radius is $M=6.75\times10^{11}M_\odot$.
The solid lines in all four panels of Fig.~\ref{small3_std} represent the standard 
reference halo. 

For comparison, we generate halos with masses above and below that of the reference 
halo. To generate these, we start with the same $P(k)$, smoothed on the same $l_0$ 
scale as before, but choose peaks of different heights, or $\sigma$'s in the initial 
density fluctuation field. An initial $2\,\sigma$ peak collapses to a halo of final 
virialized mass $M=3.5\times10^{10}M_\odot$, and a $M=2.7\times10^{12}M_\odot$ halo 
forms from a $4\,\sigma$ peak. Higher peaks in the initial density field start 
collapsing earlier, and end up as more massive and more concentrated halos. 
This trend can be seen in the upper left panel of Fig.~\ref{small3_std}. Because
our halos are isolated (i.e. their formation does not involve major mergers
which can change the halo's formation history in a stochastic way)
there exists a one-to-one relation (for a fixed smoothing scale) between the height 
of the primordial peak, the epoch of collapse, the virial mass and the central
concentration of halos produced in our formalism. We will discuss the origin of 
these trends in Section~\ref{imprint}.

Note that our halos have flat density cores at center: in the
case of the $M=6.75\times10^{11}M_\odot$ reference halo the core is very small,
$\sim10^{-3}\,R_{vir}$. For the $M=3.5\times10^{10}M_\odot$ the core is
$\sim 0.025\,R_{vir}$, however in both cases the physical extent is roughly
the same, $\sim 1$ kpc, smaller than are seen in most dwarf galaxies.

The upper right panel of Fig.~\ref{small3_std} plots 
density profiles divided by a density profile with a slope of $\alpha=2$,
the slope that would result in a flat rotation curve. The part of the halo 
where the density profile has slope $\alpha=2$ is flat in this plot.
The differences between NFW of concentration parameter $c=10$ (dot-dash curve) 
and our halos is clearly visible: our halo's slope changes slowly within the 
virial radius, whereas the NFW profiles change slope rapidly from $\alpha=1$ 
to $\alpha=3$ at the characteristic radius, $r_s=R_{vir}/c$.

The lower right panel casts the results in the form of the rotation curves,
with the horizontal axis plotted linearly to facilitate comparison with 
observations. More massive galaxy halos tend to be more centrally concentrated, 
and have flatter rotation curves. Less massive halos are less concentrated, and 
have slowly rising rotation curves. In general our rotation curves tend to be 
approximately flat, because the density profile slope
is close to $\alpha=2$ in the region between 0.1 and 1 $R_{vir}$. In contrast,
NFW rotation curves rise very steeply, a property which requires NFW fits to 
dwarf galaxy rotation curves to have very low concentration parameters (see 
van den Bosch \& Swaters 2001), in fact, too low compared to the N-body predictions.

The lower left panel shows histograms of the distribution of specific angular 
momentum (SAM) of dark matter within the virial radius. In principle, there are
two sources of angular momentum of collisionless dark matter: (i) bulk 
streaming motions, and (ii) random tangential motions.
The first arises due to tidal torques experienced by proto-halos, 
and is usually quantified as a dimensionless spin parameter $\lambda$ (Peebles 1969). 
Our halos do not have angular momentum of Type (i), only of Type (ii), so the net 
{\it vector} sum of angular momentum in our halos is zero. Because dark matter
is assumed to be collisionless, it is the sum of the {\it magnitudes} of angular 
momenta of particles that is relevant for the dynamics, not the vector sum. 
In all the following discussions angular momentum will mean Type (ii) angular
momentum.

The two smooth curves in the lower left panel of Fig.~\ref{small3_std} are 
SAM distributions of dark matter halos in N-body simulations, taken 
from van den Bosch \etal (2002), who present properties of 20 numerically 
simulated halos. This angular momentum is the sum total of streaming and
random motions of dark matter particles, i.e. Type (i) and (ii) combined.
Most of the SAM distributions taken from van den Bosch \etal tend to resemble 
the steeper of the two curves shown in Fig.~\ref{small3_std}, halo $\# 170$. 
Halo $\# 081$ has the shallowest of their SAM distributions. The plot shows 
that most halos generated using our technique have more angular momentum than 
typical N-body halos, even without the inclusion of Type (i) angular momentum
from tidal torques. This is consistent with our halos 
having more extended mass distributions than NFW halos.
We will return to this issue of angular momentum again in Section~\ref{comparison}.

\begin{figure} 
\plotone{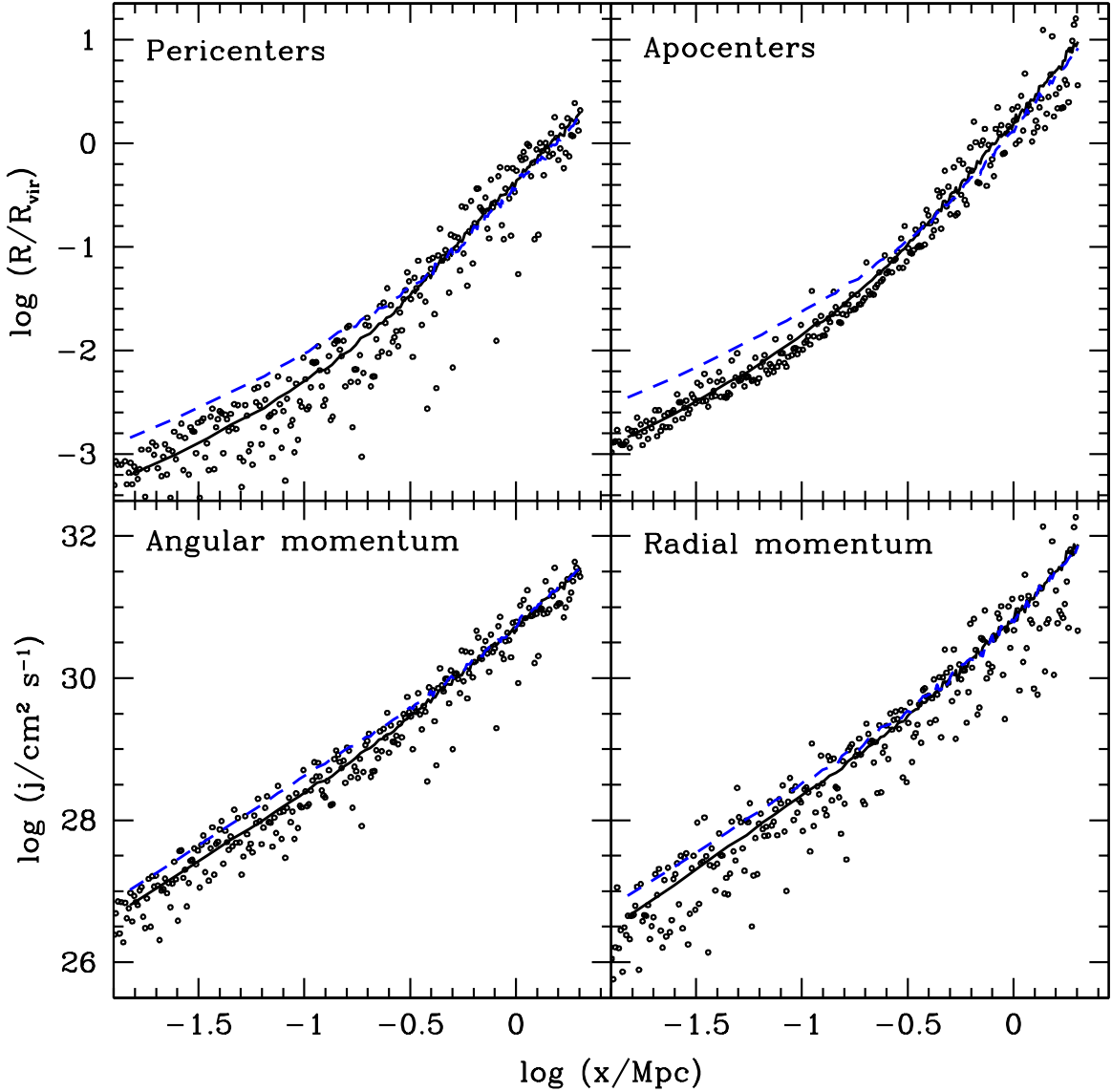} 
\vskip-0.8truecm 
\caption{Top panels shows apo- and pericenters, while bottom panels show 
angular and radial momenta of individual shells (solids dots), and averaged 
over shells from many halos at the same initial comoving position, $x$ (lines).  
Solid lines and points represent the reference halo, Section~\ref{std}. 
The dashed lines represent halos generated using a smoothed power spectrum 
with $r_f=0.1$ Mpc (Section~\ref{filter}). 
Notice the scatter in properties from shell to shell.  
Also note that shells in halos with less fluctuation power on small scales
(dashed lines) have more angular and radial momentum and do 
not penetrate deep into halo's interior (see Section~\ref{imprint}.)} 
\label{smwhysr34sj12} 
\end{figure}

The properties of individual shells in virialized reference halos are presented in 
Fig.~\ref{smwhysr34sj12}. The solid dots in top panels are the peri- and apocenters
of shells, plotted as a function of their initial, pre-collapse, comoving 
location in the halo, $x$. As expected, there is significant scatter in properties 
of individual particles, since their $\Delta {\bf v}$'s were picked randomly from a 
distribution. The solid lines are averages over shells in many halos. 
(Dashed lines represent halos considered later, in Section~\ref{filter}.) 
The average orbit eccentricity is approximately constant within the 
virialized halo, changing only from $\langle r_{peri}/r_{apo}\rangle \approx 0.3$
at $R_{vir}$ to $0.35$ at $0.01\,R_{vir}$, as orbits become more circular 
close to center. The constancy of $r_{peri}/r_{apo}$
was predicted by Nusser (2001) for halos with adiabatically varying potentials,
and power law radial profiles. It is not entirely surprising that our method, which 
conserves adiabatic invariants, but does not insist on power laws density profiles,
yields roughly constant $\langle r_{peri}/r_{apo}\rangle$ ratios. It is more 
surprising that very high resolution N-body simulations described in 
Ghigna \etal (1998) also produce constant $\langle r_{peri}/r_{apo}\rangle$
($\approx 0.2$) ratios of dark matter particles in virialized
halos. This constancy of $\langle r_{peri}/r_{apo}\rangle$ throughout the halo 
could suggest that the adiabatic approximation is largely valid for halos 
generated in N-body simulations (see also Jesseit \etal 2002).

The solid points in the lower panels of Fig.~\ref{smwhysr34sj12} plot the angular 
and radial momenta of individual particles in virialized halos. Each particle's
$j_\theta$ and $j_r$ are assumed to be constant throughout the collapse.
The radial momentum, $j_r$ and angular momentum $j_\theta$ are nearly equal
The solid lines are averages of these momenta, taken over shells in many halos. 

Before we go on, we note that the reference halo described in this section
would have looked somewhat different had we used the original RG87 method, 
which applies the rms value of random velocities, $\Delta {\bf v_{\rm rms}}(x)$
to particles in each shell. Using a {\it distribution} of $\Delta {\bf v}$ at every
$x$, as we do here, means that some fraction of particles acquire very small 
$\Delta {\bf v}$'s and $j_\theta$'s, and hence have nearly radial orbits.
These orbits steepen the density profiles somewhat, especially in the inner
halo. However, there is almost no change in density profile at larger radii.

\subsection{Comparison with NFW profiles}
\label{comparison}

Figure~\ref{small3_std} shows that halos generated using our method are
different in character from the profiles predicted by numerical simulations,
like those of NFW. This result is not surprising since the types of evolution 
that numerical N-body and our halos undergo are rather different: the former
are produced as a cumulative result of many minor and major mergers of smaller 
sub-halos, while the latter are the product of quiescent accretion of lumpy 
matter onto the primary halo. The difference between the final virialized 
halos is best illustrated by the top two panels: within
the virial radius the log-log density slope changes gradually from
around $\alpha\sim 0$ to $\alpha\sim 2$. NFW profiles, on the other
hand, have a characteristic scale-length, equal to $0.1\,R_{vir}$ in this case,
beyond which the density profile slope steepens, so that
much of the mass is piled up within 10\% of the virial radius.
This pile up is also reflected in the differences of the two angular
momentum distributions, shown in the lower left panel of Fig.~\ref{small3_std}.
The density and SAM distribution plots suggest that compared to NFW, 
the parts of our halos that lie between $\sim 0.1$ and $\sim 1\,R_{vir}$ 
retain more angular momentum.

\begin{figure} 
\plotone{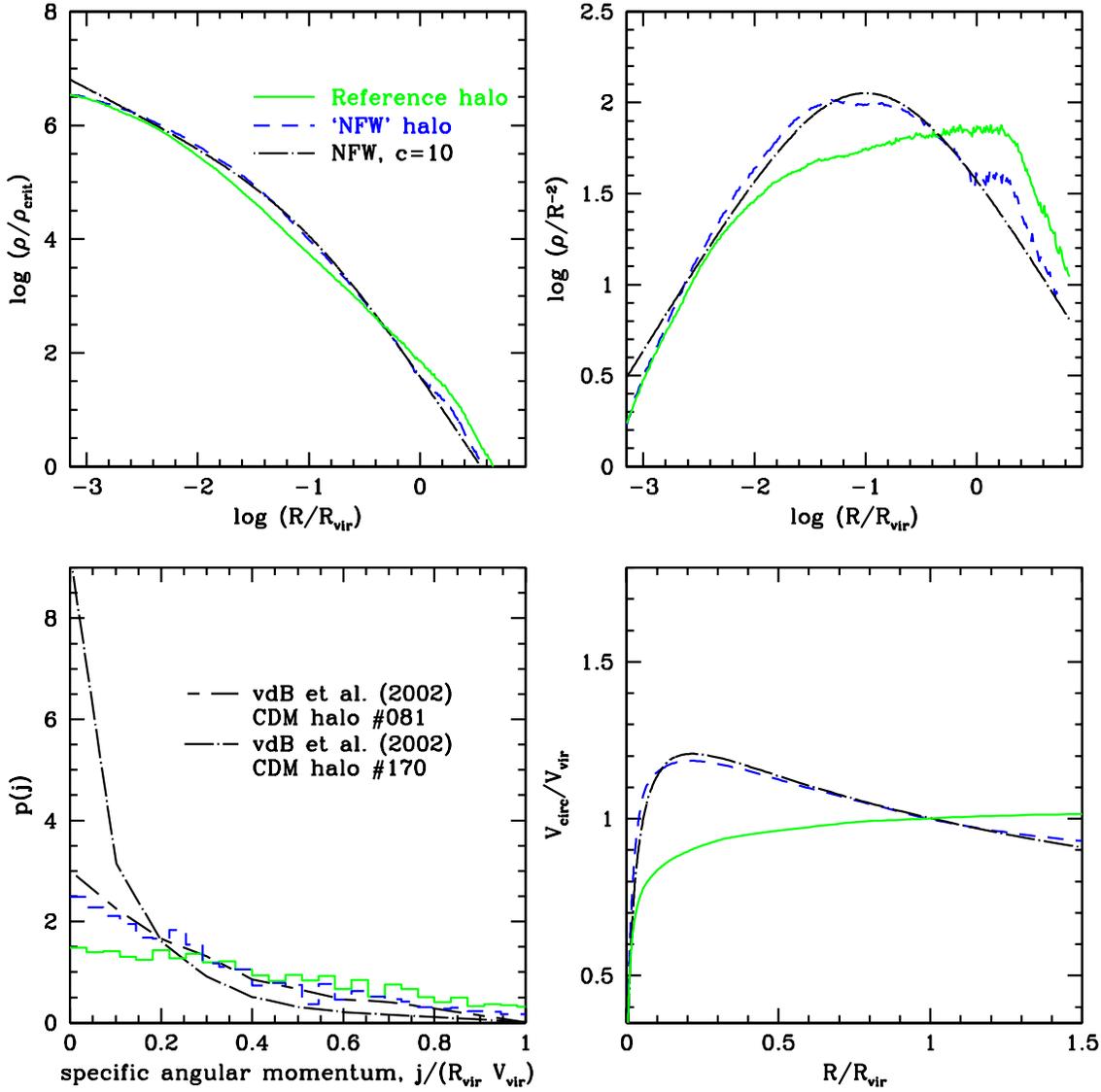} 
\vskip-0.8truecm 
\caption{Same as Fig.~\ref{small3_std}, but here we attempt to reproduce
a $c=10$ NFW halo. Starting with our reference halo (Section~\ref{std}) we 
reduced the amplitudes of the random velocities of particles that ended up 
roughly between 0.1 and 1 $R_{vir}$. The SAM distribution of this halo 
(dashed histogram in the lower left panel) is similar to that of N-body halo
\#$\,081$ (van den Bosch \etal 2002).
See Section~\ref{comparison} for details.  
}
\label{small3_NFW} 
\end{figure}

To further explore the relation between our halos and NFW we perform
the following experiment. We attempt to artificially modify the conditions 
within our halos, such that resulting halos resemble NFW. 
Figure~\ref{small3_NFW} shows two halos: the solid line is the reference
halo, while the dashed line represents a halo that was obtained by using
same parameters as for the reference halo with one exception. For shells
with initial comoving radii $x=0.35\rightarrow 1.3$~Mpc we have
arbitrarily reduced the amplitude of all random velocities, $|\Delta {\bf v}|$
by a factor of 2. How does the initial comoving $x=0.35\rightarrow 1.3$~Mpc 
radius range translate into the present day particle location within the 
virialised halo? A shell that started out in the primordial halo at a
comoving radius of $x=0.35$ Mpc has settled to a present day configuration
with apo- and pericenters at $\sim 0.033$ Mpc and $\sim 0.012$ Mpc, respectively.
Similarly, a shell originally at a comoving radius of $x=1.3$ Mpc has at
the present epoch apo- and pericenters roughly equal to 0.7 Mpc and 0.18 Mpc.
The $x=0.35\rightarrow 1.3$~Mpc range, and the factor of 2 reduction in
random velocities were selected by trial and error.
Note that we did not aim to
reproduce an NFW profile {\it exactly}; the key feature that we wanted
to emulate was the location of the break in the density profile, which occurs 
where $\alpha\approx 2$, at $r\;\simlt\; 0.1 R_{vir}$ in the NFW halo.

The corresponding change in angular momentum distribution is seen in the lower 
left panel of Fig.~\ref{small3_NFW}. The SAM distribution of the NFW-like halo
(dashed histogram) is more centrally concentrated than the SAM distribution
of the reference halo (solid histogram), and is closer to those of typical halos 
emerging from numerical simulations (smooth lines). 

The experiment that produced NFW-like halo suggests that compared to an 
isolated halo undergoing only quiescent accretion and minor mergers 
(like our halos), halos in N-body simulations lose a considerable amount of 
angular momentum between 0.1 and 1 $R_{vir}$. Since virialization 
proceeds from inside out, this means that the angular momentum loss takes 
place during the later stages of the halos' evolution, rather then very early 
on. In fact, several authors have noted that N-body generated dark matter halos 
have too little angular momentum compared to the halos of real disk galaxies, 
possibly because it was lost during repeated collisions through dynamical 
friction or other mechanisms (van den Bosch \etal 2002; Navarro \& Steinmetz 2000). 
This problem is sometimes referred to as the angular momentum catastrophe.
We suggest that halos formed via our method are closer in character to 
the real observed halos of spiral and dwarf galaxies, than are the 
N-body generated halos. 
The fact that thin extended spiral disks appear to be common in the 
Universe further supports our hypothesis, as it implies that quiescent 
formation  histories are realistic, at least for some galaxies.

\subsection{Tilting the power spectrum}
\label{tilt}

A purely scale-invariant Harrison-Zel'dovich matter spectrum has the shape
$P(k)\propto k^n$, where $k=1$. However, inflation, the most popular scenario 
for generating
scale-invariant spectra is more likely to give rise to slightly `reddish' spectra,
with $n\;\simlt\; 1$, since exponential Hubble expansion slowed down as time
advanced. Values of $n$ below 1 have been considered recently in the literature,
for example, by Zentner \& Bullock (2002), as a possible solution to the central 
density problem of galaxies. Motivated by these considerations, we tried $P(k)$ 
with $n=0.9$.  The resulting halo density profiles are only slightly shallower
than those shown in Fig.~\ref{small3_std}, so we do not show them here. 
Spectral index value smaller than $0.9$ would produce noticeably shallower density
profiles, but given the constraints on $n$ from the Cosmic Microwave Background 
(Spergel \etal 2003, Jaffe \etal 2001) and the 
clustering of Lyman-$\alpha$ clouds (Croft \& Gastanaga 1998), $n$ is unlikely 
to be smaller about $0.9$.

\subsection{Filtering the power spectrum}
\label{filter}

Another physically motivated modification of the primordial power spectrum is
filtering. Filtering the power spectrum, or, equivalently, smoothing the
density field on a given scale is meant to mimic the effect of a warm dark 
matter particle, whose mass can be related to the filtering scale, $r_f$.
Our choice of  $r_f=0.1$ Mpc is comparable to those used by other authors
(Bode \etal 2001, Knebe \etal 2002).
We make no attempt to motivate the exact choice of the value of $r_f$
starting from the mass of a hypothetical dark matter particle; instead we take 
the emperical approach and adopt filtering scales that are large enough to 
produce noticeable results, and are at the same time small enough so that 
proto-halo peaks of low mass galaxies are not completely erased.

\begin{figure}
\plotone{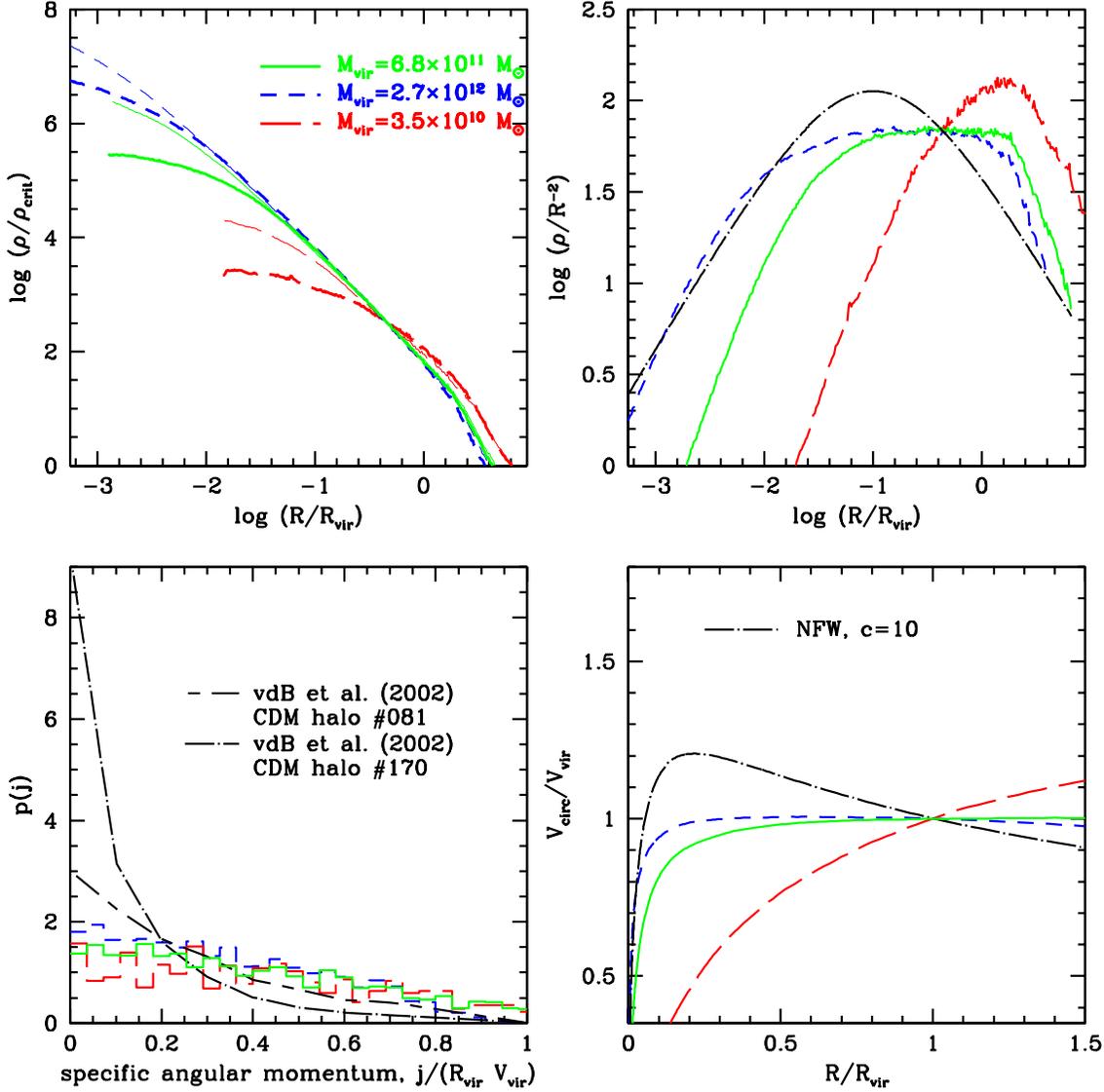}
\caption{Same as Fig.~\ref{small3_std}, but for filtered initial power spectrum:
The halos here are the same as in that figure, except that $r_f=0.1$ Mpc
has been applied to $P(k)$. The masses of the three halos are similar
to those of the corresponding halos in Fig.~\ref{small3_std}. 
For comparison, we show the three halos from Fig.~\ref{small3_std} as thin lines
in the upper left panel; the NFW profile has been omitted in this panel.
See  Section~\ref{filter} for details.}
\label{small3_filter}
\end{figure}

We smooth the power spectrum with a Gaussian filter of variable size $r_f$,
\begin{equation}
P_f(k)=P_s(k) e^{-k^2 r_f^2},
\end{equation}
and show the resulting halos in Fig.~\ref{small3_filter}. Filtering reduces 
the central densities of halos by a factor of $\simlt 10$, producing central 
cores of a few kpc in extent. Similar findings are reported by, for example,
Avila-Reese \etal (2001), who find that filtering the mass power spectra
in their N-body simulations produces halos of lower central concentration.
A quantitative explanation of why we get cores is discussed 
in Section~\ref{imprint}. The SAM distribution of all the material within 
the virial radius is very similar to that in Fig.~\ref{small3_std}, 
reflecting the fact that only the central parts of halos are affected
by filtering.

\subsection{Changing the shape of the velocity ellipsoid}
\label{LSB}

So far we have assumed that the velocity vectors arising from secondary
perturbations are randomly oriented. However, one could conceive of a process
that would not alter the magnitudes of these velocities, but orient them
preferentially tangentially with respect to the center of the halo,
thereby flattening the shape of the velocity ellipsoid.

Specifically, we do the following. Every $\Delta {\bf v}$ vector is 
decomposed into {\it two} components: the radial (with respect to the 
center of the halo) and the tangential. The latter combines $\theta$ 
and $\phi$ directions of the usual spherical coordinate system.
The tangential components of $\Delta {\bf v}$ vectors are distributed 
uniformly over the $2\pi$ radians of the tangential plane.

Let the angle between the tangential plane and $\Delta {\bf v}$ vector be
$\zeta$. To make a non-isotropic velocity ellipsoid, we keep the magnitudes 
$|\Delta {\bf v}|$ the same, but restrict the orientation of $\Delta {\bf v}$ 
to lie between angles $+\zeta_0$ and $-\zeta_0$. Within the range bounded 
by $\pm\zeta_0$ the orientations of $\Delta {\bf v}$ vectors are uniformly 
distributed. If $\zeta_0=90^\circ$ there is no restriction on the orientation
of $\Delta {\bf v}$'s, whereas if $\zeta_0=0^\circ$ only tangentially oriented
vectors are allowed. 

\begin{figure}
\plotone{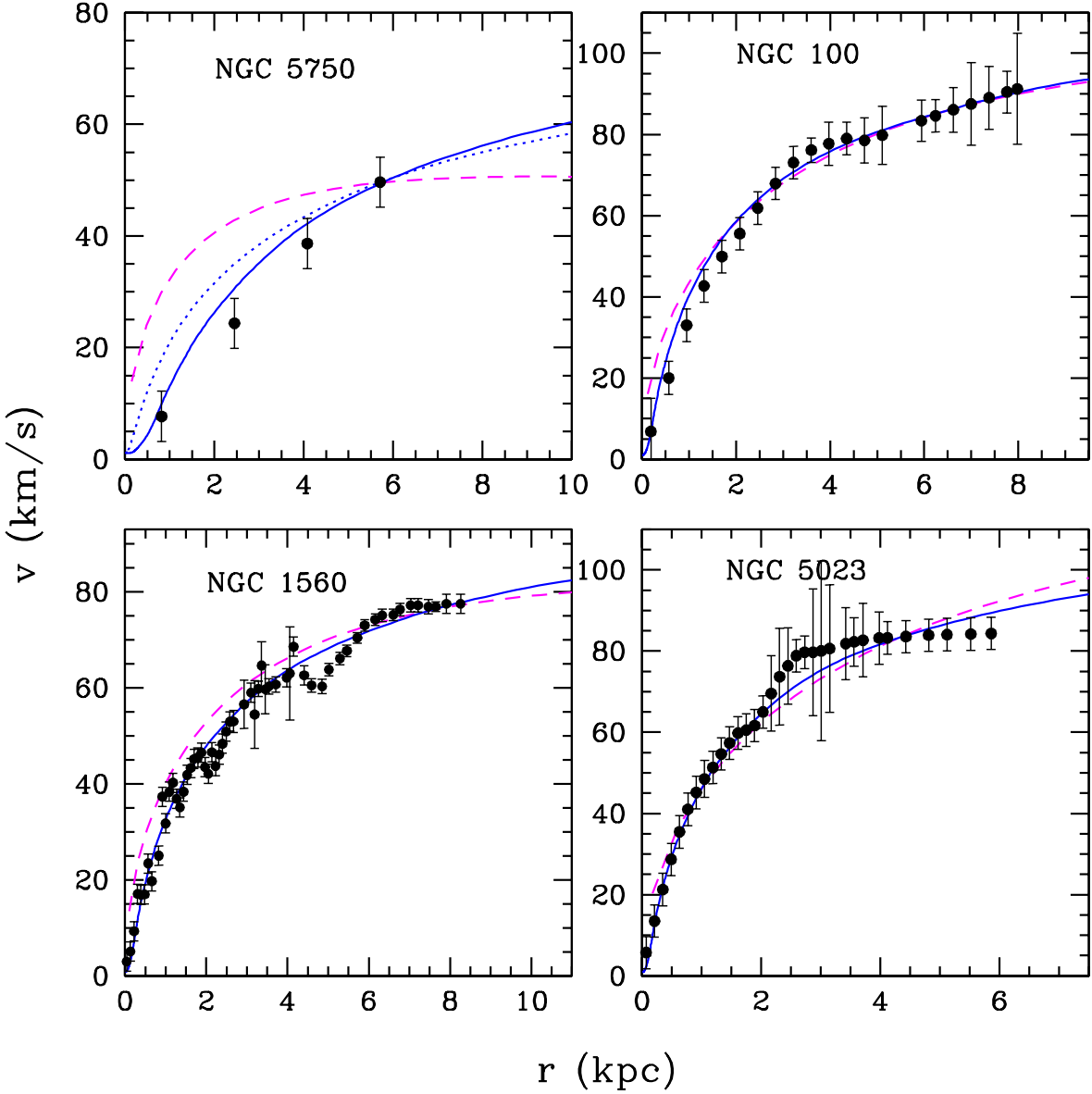}
\caption{The points with error-bars are the high-resolution rotation curves 
of LSB galaxies taken from de Blok \& Bosma (2002). The dashed lines are NFW 
fits obtained by adjusting 1 parameter: $c$ varies between 13.5 and 17 for 
these galaxies. The solid lines are our halos with tangentailly oriented 
velocity ellipsoids: $\zeta_0=20^\circ$ for NGC 5750 and NGC 100, and 
$\zeta_0=50^\circ$ for NGC 1560 and NGC 5023. The dotted curve in the upper 
left represents our halo with a spherical velocity ellipsoid, i.e. 
$\zeta_0=90^\circ$. See Section~\ref{LSB} for details.}
\label{smz4RCs}
\end{figure}

To explore the effects of a non-isotropic velocity ellipsoids we set 
$\zeta_0$ to values $<90^\circ$.
Having $\zeta_0<90^\circ$ creates an additional angular momentum barrier 
for dark matter particles, and hence results in the formation of flat 
density cores. For $\zeta_o\sim 20^\circ-50^\circ$ the final halo density 
profiles look similar to those in Fig.~\ref{small3_filter}, so we do not 
show them here. Instead, we show that these halos are a good fit to the 
rotation
curves of dark matter dominated dwarf galaxies and low surface brightness 
galaxies (LSBs).  The four LSBs in Fig.~\ref{smz4RCs} were taken from the 
recent high-resolution data of de Blok \& Bosma (2002). Of the 26 LSBs 
that are presented in that paper we picked the ones with high inclination 
angles, smooth rotation curves, and good agreement between HI, H$\alpha$, 
and optical data. 

The dashed lines in Fig.~\ref{smz4RCs} are the NFW profile fits to the last
observed point (except in the case of NGC 5023, where we fit to a point at
4 kpc). The concentration parameter $c$ was chosen to be consistent with the
NFW predictions, in other words, each rotation curve had only one free
parameter; $c$ values vary between 13.5 and 17 for these four galaxies.  
The solid lines are our halos with tangentially oriented velocity ellipsoids: 
for NGC 1560 and 5023, $\zeta_0=50^\circ$ was used, while for NGC 5750 and 100, 
more flattened velocity ellipsoids were used, with $\zeta_0=20^\circ$. 
Aside from $\zeta_0$, there is only one free parameter in our fits, the mass 
of the virialized halo. In the case of NGC 100 and 5023 ~NFW and our halos 
do equally well in reproducing the rotation curves, in NGC1560 our halo 
provides a better match to observations, and in the case of NGC5750
our halo is a considerably better fit than NFW. For NGC 5750, we also
plot the profile of a halo whose velocity ellipsoid is spherically symmetric 
($\zeta_0=90^\circ$); see dotted line in the upper left panel of Fig.~\ref{smz4RCs}.
So even a halo with spherically symmetric random velocity distribution is 
a much better fit to low mass galaxies, like NGC 5750, than an NFW profile.
Note that $r_f=0.1$ Mpc halos (Section~\ref{filter}) with randomly
oriented $\Delta {\bf v}$ vectors also provide a good fit to these LSB 
rotation curves.

\section{Understanding the halos}
\label{understanding}

\subsection{An absolute upper limit on the steepness of the central cusp}
\label{cusp1}

While the original NFW papers and later work by Power \etal (2003) showed
that the inner profile slope, $\rho\propto r^{-\alpha}$, is close to $\alpha=1$, 
Moore \etal(1998), based on their high resolution simulations argued that 
central cusps of halos are better characterized by $\alpha\approx 1.4-1.6$. 
To shed some light on this unresolved issue, we ask, 
how steep can the central density cusp get? 

First, we conducted an experiment: we arbitrarily changed the amplitude 
of random velocities imparted to dark matter particles, $|\Delta {\bf v}|$, 
by factors of 3, 10 and 100. Reducing (increasing) $|\Delta {\bf v}|$ 
amounts to reducing (increasing) the angular momentum. The results are shown 
in Fig.~\ref{smden_sv}. With each reduction of the random velocities, the 
profiles get steeper at the center. This steepening is easy to understand 
qualitatively in terms of an angular momentum barrier. 
The central density is built up by shells whose pericenters are very
close to the center of the halo. Particles with large angular momenta
are prevented from coming close to halo's center and contributing to the 
central density. This is clearly seen in the halo where $|\Delta {\bf v}|$ was 
increased by a factor of 3. Conversely, more material with low angular momentum 
results in steeper central cusps.
However, no matter how much angular momentum is reduced, it seems to be
impossible to produce profiles steeper than log-log slope of $\alpha=2$ in
the central parts of halos.
We now explain why this is the {\it absolute} steepest slope that 
can be achieved  at halos' centers.

Let us approximate the central region of the halo by
$\rho(r)=\rho_0 r^{-\alpha}$, since a single power law is a good approximation for
a limited range of radii. The potential at a point $r$ in the sphere of this
density profile is
\begin{equation}
\psi(r)=-4\pi G \rho_0 \Bigl[-{{r^{2-\alpha}}
\over{(3-\alpha)(2-\alpha)}}
 + {{{r_\infty}^{2-\alpha}}\over{(2-\alpha)}}\Bigr]
\label{haha1}
\end{equation}
where $r_{\infty}$ is the truncation radius of the halo. In this potential
a particle's radial velocity is given by
eq.~\ref{vrad}:
\begin{equation}
v_{rad}=[2(E-\psi(r))-(j_\theta/r)^2]^{1/2}
\label{haha2}
\end{equation}
We now make two approximations in this expression for $v_{rad}$. 
First, the particle's total energy 
$E$ can be assumed to be equal to its potential energy at apocenter. Its
kinetic energy can be ignored because $v_{rad}=0$, and $v_{tan}$ attains its 
minimum value at this radius. So except for the nearly circular orbits, the total energy 
is $E=\psi(r_{apo})$.  Second, for elongated orbits ~$j_\theta/r=v_{tan}$ is 
almost always smaller than $v_{rad}$ (except at apo- and pericenters), 
because the latter includes random as well as bulk motions, while the former 
includes only random motions. Therefore we can assume that $j_\theta$ in 
eq.~\ref{haha2} is zero in most cases of interest, and in particular for those
orbits which pass through the central cusp.
 Eq.~\ref{haha2} then becomes:
\begin{equation}
v_{rad}^2=2[\psi(r_{apo})-\psi(r)]=
8\pi G \rho_0 \Bigl[{{{r_{apo}}^{2-\alpha}-r^{2-\alpha}}
\over{(3-\alpha)(2-\alpha)}}\Bigr].
\label{haha3}
\end{equation}
The solid lines in Fig.~\ref{smslope2} implement Eq.~\ref{haha3}: they show 
how the radial velocity changes 
with the radial position, $(r/r_{apo})$ of the particle, for particles living 
in halos with different density profile slopes:  
$\alpha=1.99, 1.9, 1.75, ....\; 0.5$ (top to bottom). In general, $v_{rad}$ 
increases as particles get closer to center, but then becomes nearly constant 
at very small radii. 

For a single shell, the time spent within $dr$ of $r$ is inversely 
proportional to the shell's (or particle's) radial velocity. So the density 
a shell contributes at $r$ is proportional to $(r^2\;v_{rad})^{-1}$. 
If the halo mass density at $r$ is the sum of 
contributions of all shells that pass through $r$, and each shell is 
characterized by $v_{rad}$, then we must insist that 
$\rho(r)=\rho_0 r^{-\alpha}\propto (r^2\;v_{rad})^{-1}$, and so
$v_{rad}\propto r^{\alpha-2}$. When this condition is applied to solid 
curves in Fig.~\ref{smslope2} the only viable solutions that remain are 
marked by solid dots (one solution for each value of $\alpha$).
One readily sees that as the profile steepens towards $\alpha=2$ 
the orbit shapes have to get progressively more radial. To achieve $\alpha=1.9$,
$r/r_{peri}$ must be nearly 100, and so $r_{apo}/r_{peri}$ must exceed 100.
Such radial orbits are hard to achieve since even the slightest amount of 
angular momentum will prevent these extreme orbit elongations. To attain 
$\alpha=2$, $r_{apo}/r_{peri}$ of orbits must be nearly infinite. Thus there 
is an absolute upper limit on the steepness of the density cusps in real halos.

This mechanism can be understood intuitively: as particles get close
to their pericenters, their velocities increase, and so they spend less time 
close to the halo's center, and contribute little to central mass density. 
In other words, steepening up of the central cusp is a self-limiting process.

One may wonder why the mechanism just described does not impose an upper limit
on the slope of density profiles far from center. The key difference is that at
the center the density is built up by inner parts  (pericenters) of outer
shells alone. At positions in the halo other than the center, density is built
up by a combination of apocenters of inner shells, pericenters of outer shells, 
and intermediate radii shells, so the situation is more complicated.

\subsection{A practical upper limit on the steepness of the central cusp}
\label{cusp2}

While it is 
still uncertain whether the central cusp of numerically simulated 
halos is better approximated by $\rho\propto r^{-1}$~ or ~$\rho\propto r^{-1.5}$, 
cusps steeper than $\alpha\approx 1.5$ are not typically seen in simulations. 
The model presented in Section~\ref{cusp1} offers a possible explanation for this,
i.e. it allows us to place a {\it realistic} upper limit on the steepness of cusps. 

First, we compare the predictions of our analytic model, represented by solid 
dots in Fig.~\ref{smslope2}, to the data from the dark matter halos generated 
both with our semi-analytic method
and with N-body techniques. The horizontal arrows in Fig.~\ref{smslope2}
represent data taken from our dark matter halos; the cross at (-0.70, 0.05) 
represents high resolution N-body halo of Ghigna \etal (1998). These
data points were plotted as follows: the horizontal coordinate is 
log$(r_{peri}/r_{apo})$ of orbits located around $0.01-0.05\; R_{vir}$, i.e. 
in the inner parts of halos.  Log$(r_{peri}/r_{apo})$ is the minimum value that
log$(r/r_{apo})$ can attain, so the arrow symbols signify that we are plotting 
lower limits on log$(r/r_{apo})$. The vertical axis in Fig.~\ref{smslope2}
calls for the radial velocity
of the particles near the central cusp. However, our model predicts a monotonic 
one-to-one correspondence between log$(v_{rad})$ and $\alpha$,
represented by the dashed line in Fig.~\ref{smslope2}. So rather than measuring
$v_{rad}$ directly, we take the measured inner slope $\alpha$ and derive the
value of $v_{rad}$ that simultaneously solves the constraints given by 
Eq.~\ref{haha3} and by $v_{rad}\propto r^{\alpha-2}$. 

The model predictions agree rather well with the `observed' data points, and both 
indicate that steeper inner slopes require progressively more 
radial orbits. Fig.~\ref{smslope2} indicates that there is a `break' in this 
relation, at around $\alpha\sim 1.6$.
Now, radial orbits imply that the particles have small angular momenta.
In a realistic setting, any number of
numerical effects, ranging from sensitivity of equations of motion to 
round-off errors, the stiffness of the integrators, to the discreteness 
effects such as two-body heating, can increase the angular momenta of 
particles that otherwise would be on highly radial orbits. This increase in 
$j_\theta$ will lead to an increase in log$(r/r_{apo})$. For $\alpha\;\simgt\; 1.6$,
log$(r/r_{apo})$ is a very sensitive function of $\alpha$,
therefore even a small additional $j_\theta$ will quickly bring $\alpha$ to 
$\sim 1.5-1.7$, and hence lead to the flattening of the cusps. 
This can explain why cusps steeper than $\alpha\sim 1.6$ should be rare. 
To flatten the cusps beyond $\alpha\sim 1.5-1.7$ would require considerably rounder
orbits, and so larger $j_\theta$.

\subsection{Why are the halo density profiles concave?}
\label{concave}

\begin{figure}
\plotone{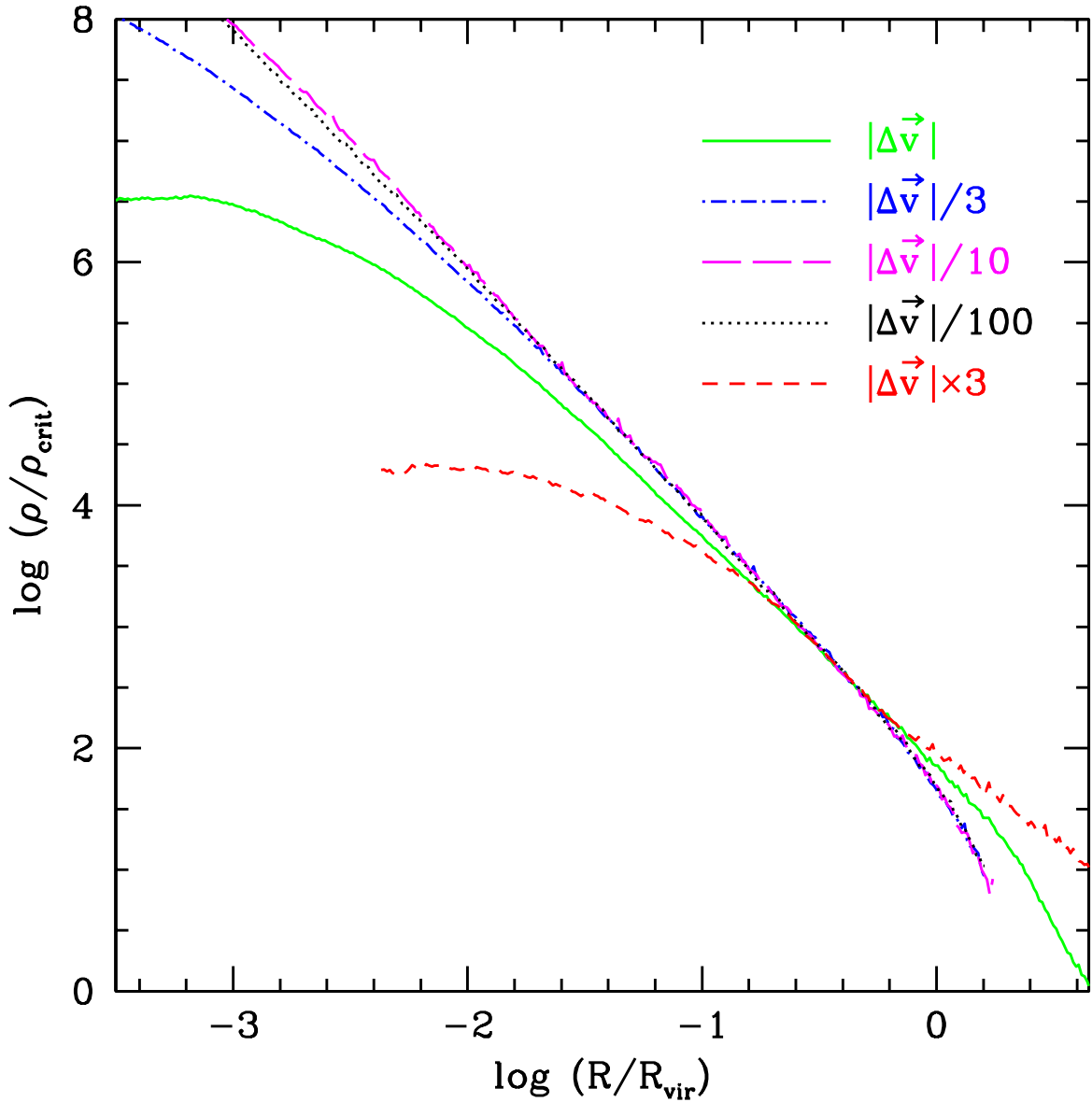}
\caption{Density profiles of halos whose random particle velocities
have been reduced by factors of 3 (dot-dash), 10 (long dash) and 100 (dotted),
and increased by a factor of 3 (short dash),
compared to the reference halo (solid line). Reducing random velocities 
means that the angular momentum of dark matter particles is reduced, which 
results in steeper central density slopes. The limiting slope,
$\alpha=2$ is discussed in Section~\ref{cusp1}, and the progressive
shallowing of the slopes in Section~\ref{concave}.}
\label{smden_sv}
\end{figure}

\begin{figure}
\epsscale{0.9}
\plotone{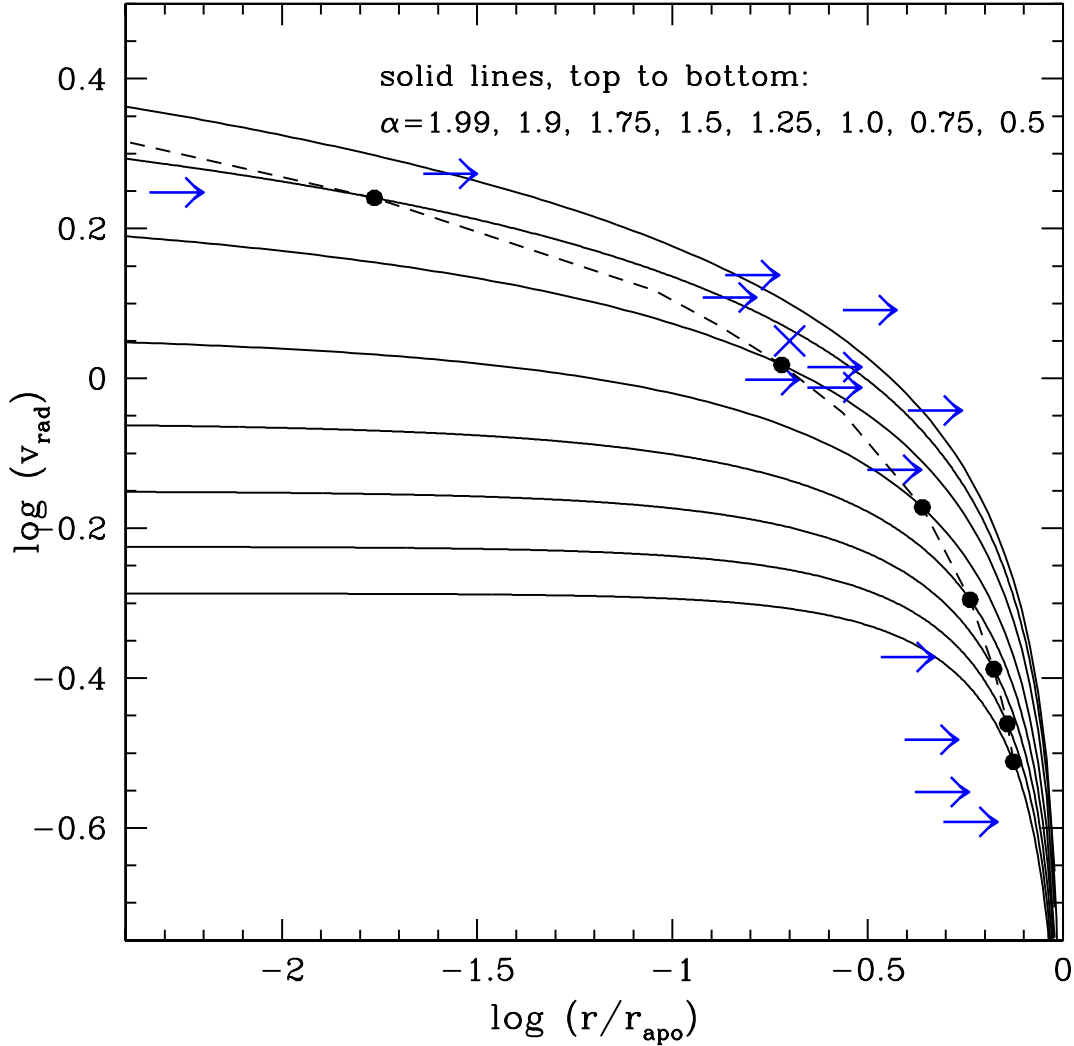}
\vskip-0.5truecm
\caption{Radial velocity vs. radial position of particles (or shells) in dark 
matter halos. The solid lines are solutions of Eq.~\ref{haha3} for a range of halo 
inner slope values $\alpha$. The solid dots connected by a dashed line are the 
predictions of an analytic model: the points satisfy both Eq.~\ref{haha3} and
the condition $v_{rad}\propto r^{\alpha-2}$; see Section~\ref{cusp1} for details.
Horizontal arrows represent data taken from our dark matter halos; 
the cross symbol at (-0.70, 0.05) represents high resolution N-body 
halo of Ghigna \etal (1998); see Section~\ref{cusp2} for more details.}
\label{smslope2}
\epsscale{1.0}
\end{figure}

Figure~\ref{smden_sv} demonstrates why, in general, density profiles tend 
to be `concave', that is, the log-log slope of density profiles steepens
monotonically with distance away from center.
First, consider a halo with negligible random velocities (dotted line in 
Fig.~\ref{smden_sv}). Such halos have slopes $\alpha\approx 2$ over 2-3 decades 
in radius. The orbits of particles are quite radial, with material being 
able to reach almost to the center. Now imagine imparting a small amount of 
angular momentum to the shells' particles. The pericenter radii are very
sensitive to $j_\theta$. Increasing $j_\theta$ by a roughly constant factor 
increases the pericenters of most shells by a constant factor as well.
This withdrawal of material has a disproportionately larger effect on the 
smaller volume inner regions of the halo, producing concave profile shapes.  
Increasing $j_\theta$ results in progressive shallowing of the slope towards 
the center of the halos. Figure~\ref{smden_sv} illustrates this steady 
shallowing: the increase in $|\Delta {\bf v}|$, and hence $j_\theta$ 
has the greatest impact on the inner density slope.

The limiting case of $\alpha=2$ slope attained for pure radial motions has 
already been noticed in the literature, in particular by Avila-Reese \etal (1998)
and Subramanian \etal (1999). These authors also note that introduction of
angular momentum results in the shallowing of the central parts of halos. 
Avila-Reese et al. (2001) attain soft cores in halos after introducing 
significant thermal velocities. These results are in full agreement with our 
work. When we increase the amplitude of random motions, and hence angular 
momentum, the resulting halos develop large flat density cores, as illustrated 
by the short-dash line in Fig.~\ref{smden_sv}. 

\subsection{A universal relation between $r_{peri}/r_{apo}$ and $j_\theta/j_r$}
\label{Jcurve}

\begin{figure}
\vskip-0.9truecm
\plotone{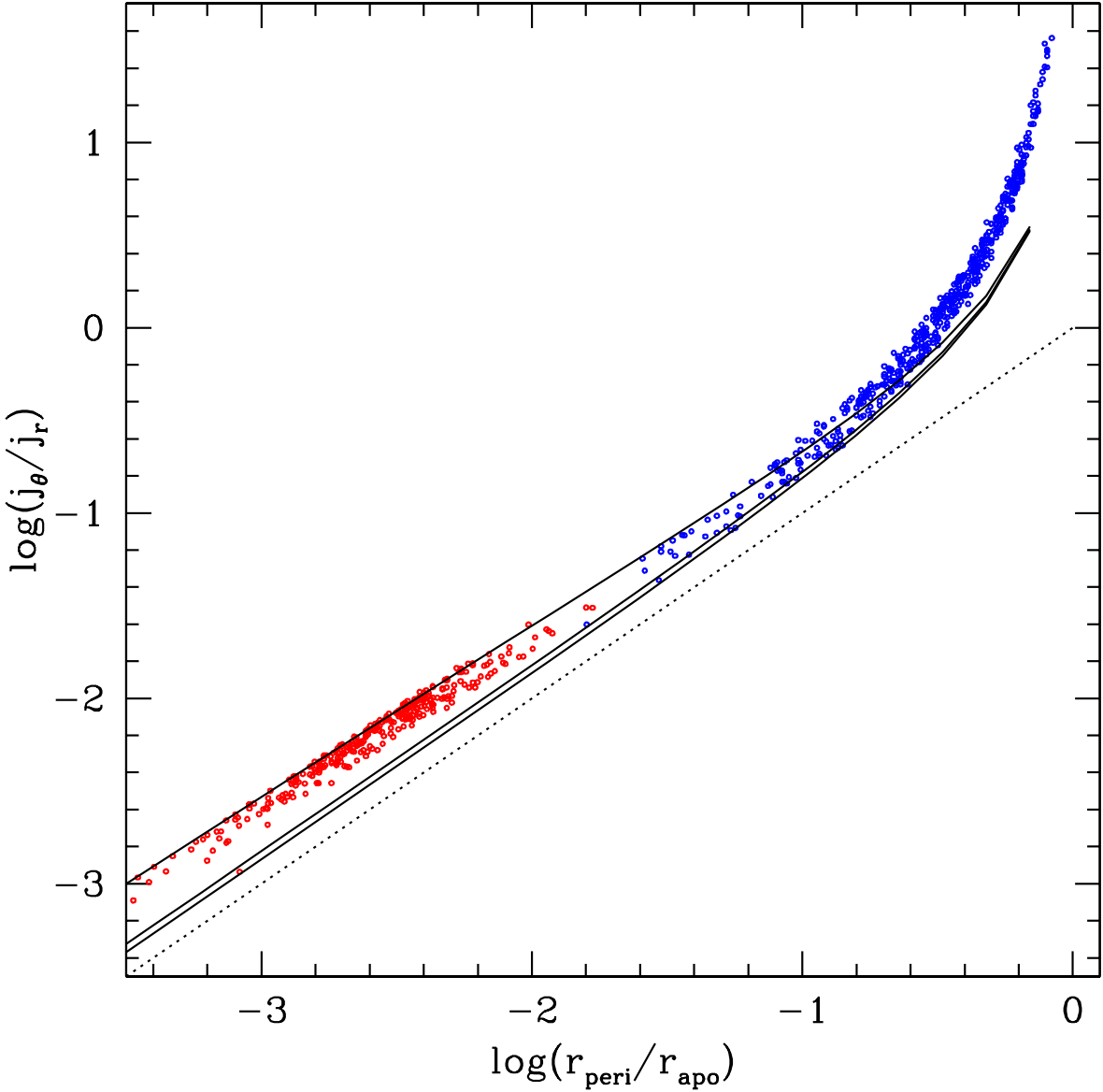}
\vskip-0.4truecm
\caption{A relation between $r_{peri}/r_{apo}$ and $j_\theta/j_r$ for shells
in our halos. In spite of the dispersion in individual shell properties
(shown in Fig.~\ref{smwhysr34sj12}) all shells in all halos obey this well 
defined relation, which can be understood in terms of the model presented 
in Section~\ref{understanding}. The three solid lines are predictions of this 
model for non-circular orbits in halos with $\alpha=1.99,$ 1, and 0.5 
(top to bottom). The dotted line is the relation $r_{peri}/r_{apo}=j_\theta/j_r$. 
Circular orbits will lie at $r_{peri}/r_{apo}=1$ and $j_\theta/j_r\rightarrow\infty$
There are two sets of points:
the upper right set belongs to the reference halo (Section~\ref{std}), 
while the lower left set belongs to the halo with almost no random motions, 
shown as the dotted line in Fig.~\ref{smden_sv}. 
See Section~\ref{Jcurve} for details.}
\label{smJcurve}
\end{figure}

Even though the values of apocenters and pericenters and of angular and radial 
momenta vary substantially from shell to shell within a given halo 
(see Fig.~\ref{smwhysr34sj12}), and even more so between different types of
halos, the relation between $r_{peri}/r_{apo}$ and $j_\theta/j_r$ is very tight
for shells in all halos. This is illustrated in Fig.~\ref{smJcurve}. The upper 
right island of points are particles in the reference halo 
(Section~\ref{std}), while the lower left set of points were drawn from a halo 
whose random velocities were reduced by a factor of 100 (dotted line in 
Fig.~\ref{smden_sv}). The dispersion in the relation is always small, typically 
a factor of 1.5. The relation is tight because, as we show below, it is determined 
by the shape of the gravitational potential of the halo, which is relatively 
smooth. We can understand the shape of this relation by considering two limits: 
orbits with small angular momenta, and orbits with large angular momenta. 

We first consider the small angular momentum limit.  As before (see Section~\ref{cusp1}), 
we approximate the density profile over a limited range of radii (between
apocenter and pericenter of a given orbit) by a power law: $\rho(r)\propto r^{-\alpha}$.
As explained in Section~\ref{cusp1}, the particle's energy is given by $\psi(r_{apo})$, 
with the potential given by eq.~\ref{haha1}.
At pericenter, $v_{rad}=0$; applying this to eq.~\ref{haha2} we get:
\begin{equation}
{{j_\theta}\over{r_{peri}}}=
\Biggl[{{8\pi G\rho_0}\over{(3-\alpha)(2-\alpha)}}\;r_{apo}^{2-\alpha}
\Bigl(1-\Bigl[{{r_{peri}}\over{r_{apo}}}\Bigr] ^{2-\alpha}\Bigr)\Biggr]^{1/2}
\label{haha4}
\end{equation}
If $j_\theta$ is not too large, then the particle's radial velocity is given by
eq.~\ref{haha3}, and its radial momentum is
\begin{equation}
j_r=\int_{r_{peri}}^{r_{apo}}\;v_{rad}\;dr=
\Biggl[{{8\pi G\rho_0}\over{(3-\alpha)(2-\alpha)}}\;r_{apo}^{2-\alpha}\Biggr]^{1/2}
\int_{r_{peri}}^{r_{apo}} 
\Bigl(1-\Bigl[{{r}\over{r_{apo}}}\Bigr]^{2-\alpha}\Bigr)^{1/2}\;dr.
\label{haha5}
\end{equation}
Combining eqs.~\ref{haha4} and ~\ref{haha5} we get,
\begin{equation}
\Bigl({{r_{peri}}\over{r_{apo}}}\Bigr)\Big/\Bigl({{j_r}\over{j_\theta}}\Bigr) =
{{r_{apo}^{-1}\int_{r_{peri}}^{r_{apo}}
\Bigl(1-\Bigl[{{r}\over{r_{apo}}}\Bigr] ^{2-\alpha}\Bigr)^{1/2}\;dr
}\Bigg/{\Bigl(1-\Bigl[{{r_{peri}}\over{r_{apo}}}\Bigr] ^{2-\alpha}\Bigr)^{1/2}}}
\label{haha6}
\end{equation}
For a fixed value of $\alpha$ of less than 2, the right hand side is roughly constant;
so $j_\theta/j_r$ is approximately proportional to $r_{peri}/r_{apo}$ with a constant of
proportionality that depends on $\alpha$. The three solid lines in 
Fig.~\ref{smJcurve} were obtained using eq.~\ref{haha6}, for 
$\alpha=1.99,$ 1 and 0.5 (starting from the top). 

Equation~\ref{haha6} breaks down for shell orbits that are close to circular, 
because we neglected $j_\theta$ in estimating $v_{rad}$. So the solid lines in 
Fig.~\ref{smJcurve} cannot continue to hold true for more circular orbits.
However, in the limit of 
circular orbits we have a simple relation: as $j_\theta/j_r$ gets large, as it 
would for orbits approaching circular shape, $r_{peri}/r_{apo}$ approaches 1, 
giving us a convergence point in the upper right corner of Fig.~\ref{smJcurve},
whose location is independent of $\alpha$. 

All the data points from our halos conform to this model.
The lower left set of points belong 
to a halo with a nearly constant $\alpha=2$ density profile, and are 
clustered just below the corresponding model line. The upper right set 
of points belong to a halo whose density profile slope varies considerably 
with distance from center, and are scattered between the curves for 
$\alpha\sim 1$ and $\sim 2$. 
In fact, particle orbit parameters of any halo considered here would also trace 
out the same curve. For example, the halo designed to mimic the NFW halo 
(Section~\ref{comparison} and Fig.~\ref{small3_NFW}) obeys this relation,
even though it was generated somewhat artificially, by removing some amount of
angular momentum from outer shells. We argue that this relation between 
$r_{peri}/r_{apo}$ and $j_\theta/j_r$ and its dependence on the slope of the 
density profile of the dark matter halo will continue to hold when the 
spherical symmetry condition is relaxed, i.e. for mildly triaxial mass 
distributions, because the derivation of the relation depends on the sphericity 
of the potential, which is always rounder than the corresponding mass 
distribution. We speculate that all {\it virialized} shells will obey 
this relation, regardless of how the halo reached virialization.

\subsection{Initial $P(k)$ and shapes of virialized halos}
\label{imprint}

From the experiments carried out in Section~\ref{tilt} and \ref{filter}
we see that if the initial $P(k)$ is modified such that it has less power on 
small scales compared to the standard power spectrum, then the final virialized 
halos will have central density profiles that are shallower compared to the 
standard case of Section~\ref{std}. For example, Fig.~\ref{small3_filter} 
illustrates that filtering the initial high-$z$ power spectrum on the comoving
scale of $0.1$~Mpc results in shallow density cores. 
We argue that these characteristics 
can be traced back to a single cause: the dependence of the turn-around radius 
and collapse time on the interior average density in the pre-collapse halo. 

The turn-around radius, $r_{max}$, and the time taken to reach $r_{max}$,
both go inversely with the average density interior to the shell 
(eqs.~\ref{rtheta} and \ref{ttheta}). The density profiles of proto-halos
are proportional to the correlation function and thus if small scale power 
in $P(k)$ is reduced, the central density of high-$z$ proto-halos is lowered. 
Consequently, inner shells are able to expand further compared to the 
standard $P(k)$ case, and it takes them longer to do so. This further implies 
that shells will collapse and virialize later.
(Note that the outermost shells 
are not affected, since the average interior density for distant shells is 
hardly different from that of the standard case.)
We now discuss specific consequences of this.\\
\indent{\hskip-10pt}{\bf (i) Displacement of particles:}
Since the inner shells now take longer to expand, secondary perturbations 
have a longer time to act on the dark matter particles. However, the perturbation
power spectrum is truncated at short scales. These two effects partly cancel
each other, but still result in a net increase in the final displacement of 
particles (eq.~\ref{Fr}).\\
\indent{\hskip-10pt}{\bf (ii) Random velocities of particles:}
The magnitude of the final rms random velocity, $|\Delta {\bf v_{\rm rms}}(x)|$ 
of eq.~\ref{deltav} is a product of the velocity growth 
factor, $F_v$, which describes the time evolution of random velocities, 
and the rms value of the initial acceleration, arising from secondary 
perturbations, eq.~\ref{deltag0}. The velocity growth factor is increased
when there is less mass fluctuation power on small scales, because 
there is more time for the perturbations' effects to accumulate.
At the same time, the rms value of the initial acceleration is somewhat decreased 
because there is less contribution to the integral in eq.~\ref{deltag0} 
from power on small scales, due to the increase in particle displacements, 
as discussed in (i). Because of these two opposing effects, the magnitude of 
$|\Delta {\bf v_{\rm rms}}(x)|$ is only slightly increased.\\
\indent{\hskip-10pt}{\bf (iii) Angular and radial momenta of particles:} 
What matters most for the subsequent dynamics of the halo are the angular and 
radial momenta of dark matter particles in the inner shells. Despite the fact 
that $|\Delta {\bf v_{\rm rms}}(x)|$ is only somewhat increased, the momenta are 
increased significantly because of the increase in the turn-around radius, see 
eqs.~\ref{jtheta} and \ref{jr}. The bottom panels of Fig.~\ref{smwhysr34sj12}
depict the change in momenta for halos formed from 
filtered initial power spectra (dashed line), as well as from the
standard initial conditions (solid line). Increased $j_\theta$ prevents 
particles from coming close to the halo center (the upper panels of 
Fig.~\ref{smwhysr34sj12}) thereby decreasing the central density.

The qualitative explanation given above, and especially the role of $r_{max}$
and turn-around time in determining $j_\theta$, also helps understand why
more massive halos are more concentrated (Section~\ref{std}) than less massive
halos. Higher $\sigma$ 
peaks (progenitors of more massive halos) have greater density contrasts
at their center, and so shells do not expand far before beginning to collapse. 
This reduces $j_\theta$, and allows halos to become more concentrated. 
We caution that this trend of increased central concentration as a function 
of mass applies only to halos that started out as peaks in the density field
smoothed with a fixed $l_0$ scale. Our conclusions do not mean that, for
example, clusters of galaxies will be very much more centrally concentrated 
than galaxies, since different smoothing scales would apply in the two cases.

This quantitative argument explains why halos formed through quiescent
accretion of small sub-halos, with no major mergers, would retain some 
memory of the initial conditions, such as the shape of the power spectrum
(also see Ryden 1988b). 

\section{Summary and Future Work} 

In this paper we used 
an analytic technique to explore dark matter halo formation of
galaxy-mass halos. Our method is derived from Ryden \& Gunn (1987) and
assumes conservation of angular and radial momenta of halo shells, as
the halo collapses and virializes. The important feature of the method
is the inclusion of secondary perturbations, which lead to radial and
tangential velocity dispersion of halo particles. The resulting angular
momentum of the halo due to random tangential velocities
makes collapse proceed in a fashion very different from the purely 
radial collapse considered in early analytical work. In contrast to other 
recent analytical works, our halos acquire 
angular momentum from secondary perturbations, whose
magnitude is derived from the fluctuation power spectrum, 
in a self-consistent way. The slope, filtering, and normalization of 
the initial power spectrum can be specified as desired. We have improved 
upon the original RG87 formalism by allowing dark matter particles to have 
a range of random velocities, which vary in both magnitude and direction 
at any given radial location in the halo. This allows us to explore 
different halo formation scenarios which produce different velocity ellipsoids. 
As a result, we have
gained a deeper understanding of the process of halo formation.

Our galaxy-mass halos formed under standard cold matter matter initial
conditions differ from universal profiles found in Navarro \etal (NFW) 
and Moore \etal (1998) numerical N-body simulations.
While the latter have a pronounced density profile slope change at about 
$10\%$ of the virial radius, our halos exhibit a much more gradual slope 
change. Specifically, the double logarithmic density slope of our halos,
steepens to $\alpha=2$ at around the virial radius, whereas the numerical
profiles' slopes are close to $\alpha=3$ well within the virial radius. 
This difference results in our rotation curves being much flatter 
over a longer range in radii compared to the universal rotation curves.
The specific angular momentum distribution of our halos is also 
different from those of numerical simulations; the former is more extended,
consistent with the halo mass distribution being also more extended.
Under standard initial conditions our halos develop small central density cores. 
However, if the velocity ellipsoid is made somewhat tangential then the 
resulting cores are large enough to reproduce rotation curves of LSB galaxies.

All in all, our halos appear to be a closer match to the halos of spiral 
and dwarf galaxies, than are N-body halos. This may indicate that the halos
of real late-type disk galaxies undergo a formation scenario similar to the one
depicted by our method, i.e. collapse proceeds through a quiescent accretion 
of lumpy material and minor mergers, rather than through a merger-driven 
formation process characteristic of fully hierarchical models.

Guided by the properties of our virialized halos generated from standard as 
well non-standard and `experimental' initial conditions, and aided by a simple 
analytical model, we were able to answer some general questions regarding halos. 
In particular, we show that the inner cusp slope cannot exceed $\alpha=2$. 
Moreover, in any realistic setting, the cusp will be shallower than 
$\alpha\sim 1.5-1.7$. This is a promising explanation for why numerically 
simulated halos have central cusps in the range $\alpha\sim\; 1- 1.5$.
However, it is also possible that the central cusps are due to some
lingering effects of numerical noise.

In spite of the large dispersion in properties of individual halo shells 
(apo- and pericenters, radial and angular momenta), and the differences in 
shapes of halos formed from a range of initial $P(k)$ and endowed with 
various amount of random particle motions, all shells obey a tight relation 
between $r_{peri}/r_{apo}$ and $j_\theta/j_r$ ratios. We derive this relation 
analytically, and speculate that because it relies mostly on the shape of the 
halo's potential, the relation is `universal'. In other words, we expect
that particles of any virialized halo, regardless of how it was assembled, 
will obey this relation. Finally, we explain why halos generated via our
formalism, i.e. through quiescent accretion of mass,
retain some memory of initial conditions, in particular, the shape of the 
primordial fluctuation power spectrum.

Though we understand halos formed via our method relatively well, and 
have gained some insight into the numerically generated N-body halos, 
we have yet to uncover the reason for one of the main features of NFW 
halo profiles: the rather abrupt change in slope at the characteristic radius,
well within the virialized portion of the halo. A promising step in that 
direction is that we now know how to arrive at NFW profiles within our 
semi-analytical formalism. 
Compared to a quiescent accretion, numerically generated halos
must have lost a considerable amount of their angular momentum in the outer parts,
roughly between 0.1 and 1 $R_{vir}$, possibly through dynamical friction or 
other mechanisms.

As the next step in our work, we plan to extend our halo formation technique
to include global tidal torques, axisymmetric collapse, and effects of mergers. 

\acknowledgements
We would like to thank the anonymous referee for suggesting that we 
compare our halos' velocity rotation curves to those of the observed LSB 
and dwarf galaxies; this and other suggestions greatly improved the paper.
AB is supported by an NSERC Discovery Grant.  He also
acknowledges the hospitality extended to him by the Canadian
Institute for Theoretical Astrophysics during his tenure
as a CITA Senior Fellow.
JJD was partially supported through NSF grant AST-990862 and the
Alfred P.\ Sloan Foundation.


\appendix
\section{A summary of RG87 method}\label{adnauseum}
 
Here we will describe, in detail, the analytical galaxy formation
formalism developed by Ryden and Gunn (1987). Our presentation differs
somewhat from theirs, but we would like to stress that everything presented
in this Appendix is taken from RG87. We will often quote equation numbers
to help the reader make the connection with their original work. 

\subsection{Initial set-up}\label{setup}

Ryden and Gunn (1987) start with an Einstein-de Sitter $\Omega=1$ universe
model, and $H_0$ of 75 km~s$^{-1}$Mpc$^{-1}$. They assume that the matter 
density is dominated by collisionless cold dark matter, CDM.
The primordial spectrum of matter density perturbations is 
Harrison-Zel'dovich 
and the transfer function was taken from Appendix G of Bardeen \etal (1986).
The post-recombination linear fluctuation spectrum, $P(k)$, is smoothed 
on comoving scales $l_0=120$~kpc, small compared to the linearized size of 
the galaxy, i.e. size of the galaxy {\it before collapse}. 
The smoothing scale corresponds to the mass of the central core of the halo, 
$M=(4\pi/3)\rho_0 l_0^3$, and is about $10^9 M_\odot$ for the above choice of
$l_0$. The smoothed power spectrum is related to the unsmoothed one by, 
$P_s(k)=P(k) e^{-k^2 l_0^2/2}$, for a top-hat filter.
The smoothed spectrum is then normalized such that the 
{\it rms} fluctuations in mass in top-hat spheres of radius $x_0=10$~Mpc
are unity at the present time. 
The unsmoothed spectrum is not used from
now on; all relevant quantities are derived using $P_s(k)$.

\begin{figure}
\plotone{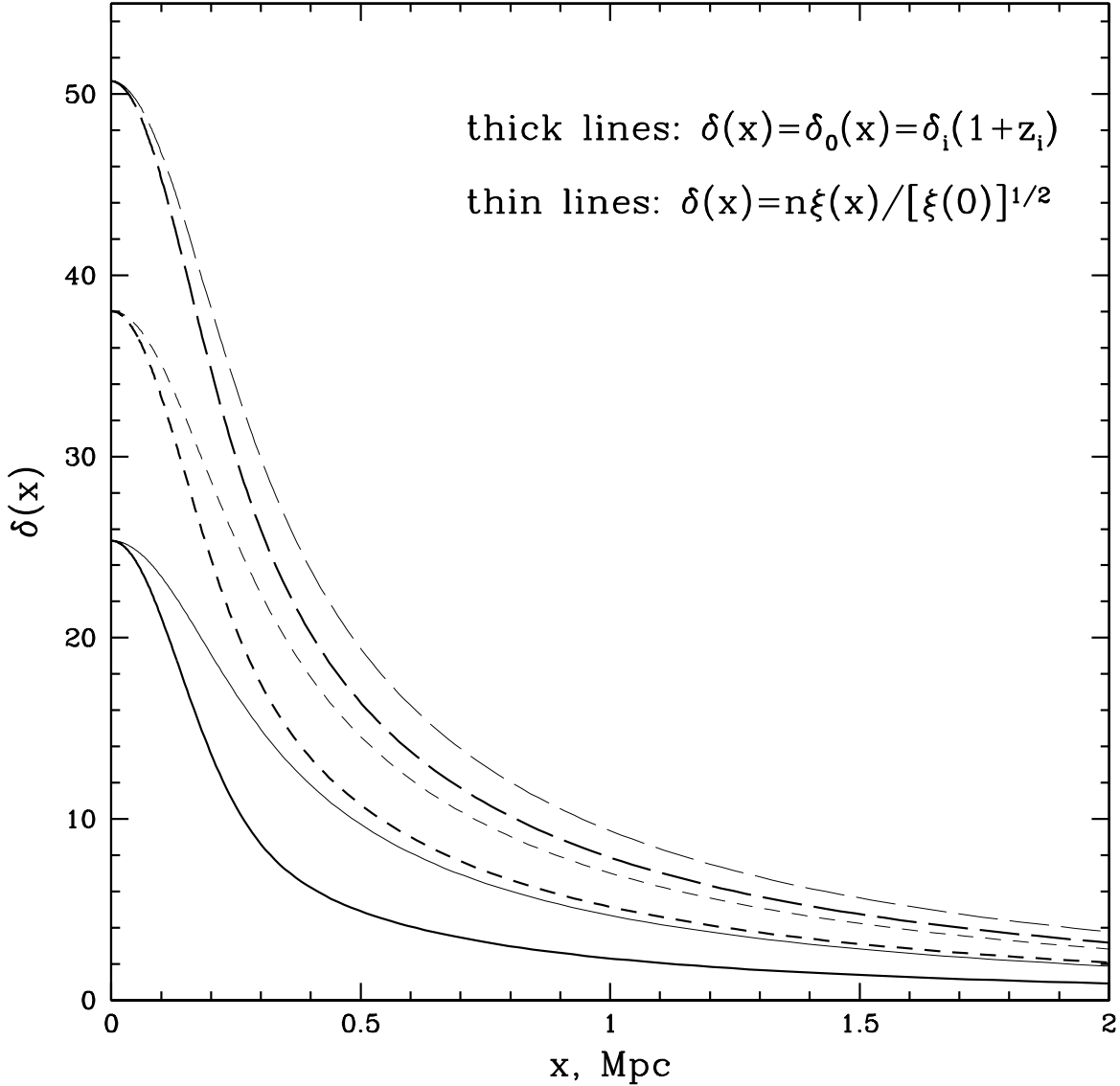}
\caption{Initial density profiles of dark matter halos (see Section~\ref{setup}).
Thin lines (the higher of each of the three sets of lines) are proportional to 
the two-point correlation function, and represent density run averaged around 
local density maxima or minima in the initial Gaussian random field. 
Solid, short dashed, and long dashed lines represent 2, 3, and 4 $\sigma$ peaks.
Thick lines are density profiles around maxima only, and are used as the initial 
profiles of proto-halos in our computations.}
\label{RG87fig2}
\end{figure}

Galaxies form from high peaks in the density field, high enough so that
they stand out above the `noise' and dominate the infall dynamics of the
surrounding matter. What is the mean
excess density distribution around a peak? Density profile around particles 
located at the local maxima {\it and} minima of the density field is given by 
$\langle{{\delta\rho}\over\rho}(x)\rangle=\langle\delta(x)\rangle=n\xi(x)/\xi(0)^{1/2}$ 
(Peebles 1984), where $\xi$ is the two-point mass correlation function, and $x$ 
is the comoving separation.
Eliminating minima, the density excess around centers of local density peaks
was derived by Bardeen \etal (1986), and is given by eq.(9) of RG87. 
The amplitude of any 
given peak is expressed in terms of its $\sigma$ deviation, where 
$\sigma=\xi(0)^{1/2}$, and $\xi(0)$ is the zero lag value of the two-point 
function.
Thus the central density contrast of an $n\sigma$ peak is $n\xi(0)^{1/2}$. 
Given that galaxies are rather common, they must have formed from 
peaks that are not very rare, say, 2-4 $\sigma$ peaks. 
Density profiles for $n=2,3,$ and $4$ 
and plotted in Figure~\ref{RG87fig2}. Thin lines are density runs around
maxima or minima, and are proportional to $\xi(x)$. Thick lines show
$\delta_0(x)$, the initial excess density distribution around a peak linearly 
evolved to the present day. The density profile of an initial, pre-collapse halo,
at early times, $\delta_i(x)$, is related to $\delta_0(x)$
by the linear growth factor: $\delta_i(x)=\delta_0(x)/(1+z_i)$. 

The reason why density profiles around peaks (thick lines in Fig.~\ref{RG87fig2})
are steeper than those around density extrema (thin lines) is easy to understand 
in a qualitative way. Around centers of peaks the density decreases with 
radius, around centers of valleys it rises. `Subtracting' the contributions of 
the valleys from the overall correlation function we get the profiles of peaks, 
which would then have density dropping off faster with $x$ than a correlation 
function. Now consider very high peaks in the
density field: a disproportionately large fraction of particles reside in these,
and because the correlation function is weighted by the square of the particle 
density the profile of the very high density peaks will be close to that of the
correlation function. 

The density profiles in Fig.~\ref{RG87fig2} are not singular at center, but
have roughly constant density cores. This is because smoothed power spectrum was
used to produce $\xi(x)$ and hence $\delta_0(x)$.

So far we have a perfectly spherically symmetric initial density profile. 
However, the initial density peak will in general have a triaxial shape, leading 
to non-spherical collapse. The dynamics of the halo collapse are dictated 
by the potential, which, being a double integral over all space, is much 
rounder than the mass distribution. Therefore the effects of intrinsic 
triaxiality of initial density peaks are smaller than those due to the 
secondary perturbations, and so can be ignored. Furthermore, initial triaxiality 
is less severe in larger, 2-4 $\sigma$ peaks (Bardeen \etal 1986), which are
the subject of the present study.
With this caveat the smooth part of our density peak is still described by 
RG87 eq.(9). The corresponding profiles (thick lines in Fig.~\ref{RG87fig2})
are the starting point of further calculations.

In addition to the smooth halo, RG87 consider contributions from the secondary 
perturbations which arise from the same Gaussian random field that gave rise
to the main halo. The effects of the secondary perturbations on the dynamics
and final profiles of halos are the main topic of their paper. 
Strictly speaking, the rms amplitude of 
secondary fluctuations is reduced close to the central peak because of the 
constrained nature of the field; for simplicity RG87 ignore this small effect.

The overall initial density profile, linearly evolved to the present day, 
can be written as,
\begin{equation}
\rho({\bf x})=\rho_0[1+\delta_0(x)][1+\epsilon_0({\bf x})],\label{den}
\end{equation}
where $\rho_0$ is the present day background density, density excess due to
the main halo is $\delta_0(x)$, and is assumed to be spherically symmetric, 
and $\epsilon_0({\bf x})$ is the density excess contributed by the random 
secondary perturbations. 

The initial set-up is now complete; next we address the dynamics of 
halo collapse. The plan is as follows.  In Sections~\ref{one} and \ref{two} 
we consider the time evolution of $\delta(x)$ and $\epsilon({\bf x})$, 
respectively, until shells reach turn-around radius; in Section~\ref{three} 
we calculate the acceleration, velocity, and displacement of the dark matter 
particles due to secondary perturbations $\epsilon({\bf x})$; and in 
Section~\ref{four} we follow the shells past the turn-around and through the 
non-linear process of shell-crossing, and describe how to arrive at the final, 
equilibrium state of the halo.

\subsection{Evolution of $\delta(x)$ until turn-around}\label{one}

Let us divide the main smooth spherically symmetric halo, $\delta_0(x)$, 
into many concentric mass shells. Each shell is uniquely labeled by
$x$, its initial comoving radius. The halo represents an upward 
departure from average background density, i.e. density interior to any 
shell is greater than critical at all times. In such cases the time evolution 
of a shell until it reaches turn-around is given by a set of parametric 
equations (Gunn \& Gott 1972, Peebles 1980),
\begin{equation}
r(\theta)={1\over 2}x{\bar\delta_0}^{-1}(1-\cos\theta)
\label{rtheta}
\end{equation}
\begin{equation}
t(\theta)={3\over 4}t_0{\bar\delta_0}^{-3/2}(\theta-\sin\theta),
\label{ttheta}
\end{equation}
where $r$ is particle's proper radius, $t$ is cosmic time, and $\bar\delta_0(x)$
is the initial average fractional density excess inside the shell 
(eq.[13] of RG87), 
\begin{equation}
\bar\delta_0(x)={3\over x^2}\; \int_0^x \delta_0(y)\;y^2\;dy,
\end{equation}
and $t_0$ is the present time. At very early times, when $\delta\approx 0$
at all radii, eqs.~\ref{rtheta} and \ref{ttheta} reduce to 
$r(t)=x(t/t_0)^{2/3}$, which means that the proper radii of shells trace the
evolution of the scale factor; at early times all shells grow at the same rate.
Much later, shells begin to reach their maximum expansion radii. A given
shell reaches its maximum radius, $r_{max}(x)$, at time $t_c(x)/2$ 
(half the collapse time) when its time parameter $\theta$ is equal to $\pi$.
Prior to turn-around, i.e. when $\theta<\pi$ the expansion proceeds according 
to the equations above, and $\theta$ and $t$ can be used interchangeably for 
any given shell. The mass of a shell, and the mass within a shell are constant; 
these relations, together with eqs.~\ref{rtheta} and \ref{ttheta} can be used
to compute the fractional overdensity for any shell, parameterized by its
$\delta_0/{\bar\delta_0}$, at a cosmic time corresponding to $\theta$, 
(eq.[23] of RG87):
\begin{equation}
\delta(\theta)+1={{9(\theta-\sin\theta)^2}\over{2(1-\cos\theta)^3}}
\Biggl(1+3\Bigl[1-{{\delta_0}\over{\bar\delta_0}}\Bigr]
\Bigl[1-{{3\sin\theta(\theta-\sin\theta)}\over{2(1-\cos\theta)^2}}\Bigr]
\Biggr)^{-1},
\label{deltatheta}
\end{equation}
This can be used to construct the density run of a halo at a fixed cosmic 
time $t/t_0$. 

\subsection{Evolution of $\epsilon({\bf x})$ until turn-around}\label{two}

As the primary density peak grows, so do the secondary perturbations
$\epsilon({\bf x})$.
The main peak is assumed to be immersed in a uniform background, therefore
the dynamics of the halo particles are dominated by the main peak.
In  particular, the growth rate of $\epsilon({\bf x},t)$ is a function of 
the local density and the local tidal field, both due to the main peak. 
This can be intuitively understood by considering two extreme cases of
secondary peaks, located in different regions of the main halo. 
First, consider a density perturbation close to the center of the halo.
The mass distribution around it is roughly spherical, so that the peak will
behave as if it were living in a Universe of a higher average density, and
hence will experience a higher growth rate. If the secondary peak is exactly 
at the center of the main peak, tidal forces vanish and the growth rate
is a function of local overdensity only, see eq.(32) of RG87.

Now consider another perturbation, located on the outskirts of the 
main peak. It will experience tidal forces arising from the very asymmetric 
distribution of mass around it: main density peak on one side, and average 
density field on the other. In a 2D or 3D situation these tidal forces would
elongate the secondary peak inducing faster collapse along the shorter axis,
leading to increased growth rate. Our case is spatially 1D, i.e. radial, 
so the tidal forces will stretch the perturbation equally in all directions, 
thereby decreasing the peak's density and reducing its growth rate.

In a statistical sense, the growth rate will depend on $\delta_0/\bar\delta_0$,
or, equivalently, $x$, either of which can be used to parameterize the strength 
of the tidal field in a spherically symmetric halo. The final expression for the 
growth rate is given by (eq.[28] of RG87)
\begin{equation}
\epsilon(x,\theta)={\epsilon_0(x)\over{\bar\delta_0}}
{{f_2(\theta)}\over{f_1(\theta)-[\delta_0(x)/\bar\delta_0(x)]f_2(\theta)}},
\label{epsilontheta}
\end{equation}
where $\epsilon_0(x)$ is the amplitude of the initial perturbation, 
given by eq.~\ref{epsilonfield}, and 
$f_1$ and $f_2$ are functions of the `time' parameter $\theta$:
$f_1(\theta)=16-16\cos\theta+\sin^2\theta-9\theta\sin\theta$, and
$f_2(\theta)=12-12\cos\theta+3\sin^2\theta-9\theta\sin\theta$.
Thus the same average growth rate is shared by all perturbations at a given 
radial $x$. The growth proceed as follows. According to eq.~\ref{epsilontheta}, 
at early times $\epsilon(t)\propto t^{2/3}$, a standard linear growth result 
for Einstein-de Sitter, and is independent of the radial location of the peak. 
Later, centrally located peaks start to grow much faster than suggested by the 
linear predictions, while the growth rate of perturbations on the outskirts of 
the halo, where $\delta_0/\bar\delta_0<0.5$, fall somewhat below the linear 
growth rate (see Fig.6 of RG87). Note that the derivation of eq.~\ref{epsilontheta} 
assumes that $\epsilon(x,\theta)$ averaged over all the secondary peaks within any 
given radius $x$ is 0, $\bar\epsilon(x)=0$; in other words secondary perturbations 
contain zero {\it net} mass. 

\subsection{Dark matter particle velocity and displacement at turn-around}
\label{three}

As the halo expands at an ever decreasing rate, the randomly placed secondary 
density peaks within the halo grow (as described in the last section) and 
exert accelerations on the dark matter particles (as described in this
section).

At any given time and proper position the random acceleration due to 
perturbation field $\epsilon({\bf r})$ is given by (RG87 eq.[40]): 
\begin{equation}
{\bf g}({\bf r},t)={\bf g}_{tot}({\bf r},t)-{\bf g}_b({\bf r},t)\approx
G\int d^3r^\prime
{{\rho_b ({\bf r^\prime},t)\epsilon({\bf r},t)}\over
{|{\bf r^\prime}-{\bf r}|^3}}({\bf r^\prime}-{\bf r}),
\label{accln}
\end{equation}
which is an integral over all space, and 
$\rho_b({\bf r},t)=\rho_0 (t) [1+\delta({\bf r},t)]$ is the background density due
to the main halo, and is related to the total density given by eq.~\ref{den} by, 
$\rho({\bf r},t)=\rho_b({\bf r},t)[1+\epsilon({\bf r},t)]$; $\rho_0 (t)$ is the
average density of the Universe at epoch $t$.

What we want to know is how this acceleration, over time, amounts to particle 
displacement, and how much extra velocity it imparts to the particles. To that 
end we need to integrate the acceleration once with respect to time to get 
velocity, and twice to get the displacement, $d({\bf r},t)$. However the current 
expression for acceleration, eq.~\ref{accln}, needs to be simplified. 

First, remembering that the density distribution in the main peak and perturbation 
field and growth rate of perturbations are functions of radial position only, we 
use $x$ instead of ${\bf x}$, and $r$ instead of ${\bf r}$.
Second, our eventual goal is the {\it rms} value of acceleration, so scalar $g$ 
will replace vector {\bf g}. 
Third, a major simplification is accomplished by decoupling the time dependence of 
acceleration, i.e. the rate of growth of acceleration, from its spatial variation.
Let the initial acceleration field due to secondary perturbations be denoted by
$g_0(x)$. Then the dimensionless rate of growth of acceleration is given by,
\begin{equation}
F_g(x,t)=g(x,t)/g_0(x)=g(r,t)/g_0(r).
\label{Fg_def}
\end{equation}
With these, the proper displacement of a particle can be evaluated as 
\begin{equation}
d_p(x,t)=\int_0^t dt_1 \int_0^{t_1} dt_2\; g(x,t_2)
=g_0(x)\int_0^t dt_1 \int_0^{t_1} dt_2\; F_g(x,t_2),
\label{Fg}
\end{equation}
and is related to the comoving displacement, $d(x,t)=d_p(x,t)/a(t)$, where $a(t)$
is the scale factor.
Next we describe the two functions, $F_g(x,t)$ and $g_0(x)$, separately.

\subsubsection {Time evolution of acceleration: $F_g(x,t)$}

In the linear growth regime in 
Einstein-de Sitter Universe model we have, using proper coordinates: 
$\rho\propto t^{-2}$, $\epsilon\propto t^{2/3}$, and $r\propto t^{2/3}$, 
so by applying these time
scalings to eq.~\ref{accln} the growth rate of acceleration becomes  
$F_g(t)\propto t^{-2/3}$, independent of radial position $x$ of the perturbation. 
This approximation is too crude for our purposes; we suspect that acceleration 
growth rate of a small patch within the halo will depend on the
local conditions at $x$ as well as time. 
In fact, to simplify the situation RG87 assume that $F_g(x,t)$ depends on the 
{\it local} conditions only, so with the help of eq.~\ref{accln} the
acceleration growth rate becomes 
(RG87 eq.[44]\footnote{Note a typo in their paper: 
$f_1(\theta)$ in the numerator of eq.(44) should be $f_2(\theta)$.})
\begin{equation}
F_g(x,\theta)={g(x,\theta)\over g_0(x)}= 
{{\rho_0[1+\delta(x,\theta)]\epsilon(x,\theta)r(x,\theta)}
\over{\rho_0\epsilon_0 x}}=
8\bar\delta_0{{f_2(\theta)}\over
{[f_1(\theta)-{{\delta_0(x)}\over{\bar\delta_0(x)}}\;f_2(\theta)]^2}},
\label{approx_accln}
\end{equation}
where expressions for $\delta(x,\theta)$, $\epsilon(x,\theta)$ and $r(x,\theta)$ 
are taken from eqs. ~\ref{rtheta}, \ref{deltatheta} and \ref{epsilontheta}. 
As mentioned earlier, for the period of time prior to turn-around we are free 
to exchange time $t$ for $\theta$; the relation between the two is fixed by the 
parametric equations \ref{rtheta} and \ref{ttheta}.

\subsubsection {Spatial distribution of secondary perturbations: $g_0(x)$}
\label{spatial_g0}

Now we turn our attention to the spatial dependence of acceleration.
We are interested in the mean square value of acceleration with respect to the 
center of the main density peak: 
\begin{equation}
[\Delta g_0({\bf x})]^2=
\Big\langle |{\bf g}_0({\bf x}_p)-{\bf g}_0({\bf x}_c)|^2\Big\rangle,
\label{deltag_diff}
\end{equation}
where ${\bf x}_p$ and ${\bf x}_c$ are the positions of particle and the
center of the halo with respect to some absolute reference frame, and
${\bf x}={\bf x}_p-{\bf x}_c$ is constant in this calculation. The average 
is over all the realizations of the Gaussian random field which describes the 
secondary perturbation field appearing in eq.~\ref{accln}:
\begin{equation}
\epsilon_0({\bf x})={1\over{(2\pi)^3}} 
\int d^3k\; \epsilon_{\bf k}\;e^{i \bf k\cdot \bf x},\quad \quad
\langle |\epsilon_{\bf k}|^2 \rangle =P_s(k)
\label{epsilonfield}
\end{equation}
A single Fourier mode ${\bf k}$ of the $\epsilon_0({\bf x})$ field will produce 
acceleration of a particle with respect to the center of the halo equal to
(see eq.~\ref{accln})
\begin{equation}
\Delta {\bf g}_{0,{\bf k}}({\bf x})
=-i4\pi G\rho_0 \epsilon_{\bf k}{{\bf k}\over k^2}
(e^{i{\bf k}\cdot{\bf x}_p}-e^{i{\bf k}\cdot{\bf x}_c}),
\label{deltag0_k}
\end{equation}
and, by symmetry, directed along the vector ${\bf k}$.
The square of the magnitude of $\Delta {\bf g}_{0,{\bf k}}({\bf x})$ is
\begin{equation}
|\Delta g_{0,{\bf k}}|^2=
(4\pi G \rho_0/k)^2 P_s(k) 2 (1-\cos {\bf k} \cdot [{\bf x}_p-{\bf x}_c]) =
(4\pi G \rho_0/k)^2 P_s(k) 2 (1-\cos {\bf k} \cdot {\bf x}),
\end{equation} 
and the dependence on the location of the origin has dropped out.
Putting all the modes together,
\begin{equation}
[\Delta g_0(x)]^2 = {1\over{(2\pi)^3}} \int |\Delta g_{0,\bf k}|^2 d^3 k,
\end{equation}
and integrating over the spherically symmetric ${\bf k}$ space we can write
eq.~\ref{deltag_diff} as
\begin{equation}
\Delta g_0(x)
=4G\rho_0 \Bigl[\int P_s(k)(1-{\sin kx\over kx}) dk \Bigr]^{1/2}.
\label{deltag0a}
\end{equation}
By symmetry, this rms acceleration with respect to the center of the halo
is a function of particle's initial comoving radius $x$, and not its 3D position 
${\bf x}$.

Equation~\ref{deltag0a} is not yet the final form of $\Delta g_0(x)$.
This expression does not discriminate between the 
different Fourier modes of $\epsilon_0({\bf x})$, allowing all the 
modes that accelerate the particle at any one instance 
to contribute to the final acceleration. The reality is somewhat different.
Perturbation modes much smaller than the displacement $d(x,t)$
of the particle from its original position, $kd>1$, are out of phase 
with the particle, i.e. they can only jiggle the particle about its position
contributing to the instantaneous acceleration,
but not amounting to any coherent displacement when integrated over time.
Since eq.~\ref{deltag0a} does not take this into account,
the short wavelength modes need to be truncated by hand by introducing an 
exponential cutoff to the power spectrum, $e^{-kd(x,t)}$.
Equation~\ref{deltag0a} now becomes (eq.[38] of RG87),
\begin{equation}
\Delta g_0(d,x)={2\over{3\pi}}t_0^{-2}\Bigl[
\int P_s(k)e^{-kd}(1-{{\sin kx}\over{kx}})dk\Bigr]^{1/2}.
\label{deltag0}
\end{equation}
(Note that we have traded the average present day density $\rho_0$ for the 
current cosmic time $t_0$, assuming Einstein-de Sitter cosmology.)

The long wavelength modes of $\epsilon_0({\bf x})$, those with $kx<1$, 
do not contribute to the acceleration of the shell with respect to 
the halo center because they displace the center of the halo and the shell 
equally. This effect has already been incorporated when deriving 
eq.~\ref{deltag0a}, the acceleration with respect to the
center of the halo. As expected these modes are suppressed, by a factor
$(1-{{\sin kx}\over{kx}})$ in eq.~\ref{deltag0a} and eq.~\ref{deltag0}.
These long wavelength modes are important for torquing the halo as a whole,
resulting in coherent rotation of the galaxy (see Ryden 1988a).

So the modes that contribute to the net acceleration are 
the intermediate wavelength modes, $d(x,t)\leq k^{-1}\leq x$.
We emphasize that for different shells, each characterized by an initial comoving 
radius $x$ and displacement $d(x,t)$, this range of relevant perturbation
modes will be different. Also note that the particle's 
displacement is assumed to be small compared to its unperturbed 
distance from the halo's center, $d(x,t)<x$, so that particles always stay
close to the shells they originate in. 

Having evaluated the spatial (eq.~\ref{deltag0}) dependence and temporal 
(eq.~\ref{approx_accln}) evolution of acceleration separately, 
we can now put them together using eq.~\ref{Fg_def}. Note that
because we are interested in accelerations with respect to the halo center,
we modify eq.~\ref{Fg_def} to read: $\Delta g(x,t)=F_g(x,t) \Delta g_0(x)$.

\subsubsection{Velocity and displacement of particles due to the 
$\epsilon({\bf x})$ field}
\label{v_and_d}

Our goal is to calculate particles' random velocities, which is the time cumulative 
effect of acceleration. However, $\Delta g_0(d,x)$ in eq.~\ref{deltag0} is a 
function of $d$ for each shell, which is not yet known. To determine $d(x,t)$
we use the fact that acceleration over time amounts to displacement, eq.~\ref{Fg}. 
We introduce a dimensionless growth factor, $F_r(x,t)$ for displacement,
related to the acceleration growth factor, $F_g(x,t)$ (eq.~\ref{approx_accln}), 
and the proper displacement by
\begin{equation}
d_p(x,t)
=\Delta g_0(x)\int_0^t dt_1 \int_0^{t_1} dt_2\; F_g(x,t)
=\Delta g_0(x)\; t_0^2\; F_r(x,t) ,
\label{Fr}
\end{equation}
The comoving displacement, $d(x,t)$, for an
initial comoving label $x$ and time $t$ is given by (eq.[37] of RG87): 
\begin{equation}
d(x,t)=d_p(x,t)/a(t)
=\Delta g_0(x)\; t_0^2\; F_r(x,t)/a(t)
\label{eqn37}
\end{equation}
We are interested in random velocities at turnaround, which is
half of the collapse time, $t=t_c(x)/2$, so we evaluate eq.~\ref{eqn37} at
that time, when $\theta=\pi$. Using eq.~\ref{ttheta} and assuming 
Einstein-de Sitter evolution of the scale factor, $a(t)=(t/t_0)^{2/3}$, we get
\begin{equation}
d(x,t_c/2)=
({4\over{3\pi}})^{2/3}\;t_0^2\; \bar\delta_0(x)\;\Delta g_0(d,x)\;F_r(x,t_c/2)
\label{eqn47}
\end{equation}
The displacement $d$ appears on both side of this equation, and thus 
can be solved for in a self-consistent fashion for every shell.  

Next, we estimate the magnitude of the extra 
velocity imparted to a typical dark matter particle
at the time of turn-around (eq.[48] of RG87),
\begin{equation}
|\Delta {\bf v_{\rm rms}}(x,t_c/2)|=F_v(x,t_c/2)\Delta g_0[d(x,t_c/2),x]t_0,
\label{deltav}
\end{equation}
where velocity growth factor $F_v(x,t)$ is defined similarly to 
$F_g(x,t)$ and $F_r(x,t)$. 
Because the secondary perturbations are located randomly within the halo,
the direction of $\Delta {\bf v}$ at any given point is random, therefore 
statistically, i.e. averaging over all the particles in any given shell, 
the squared magnitude of $\Delta {\bf v_{\rm rms}}$ should be divided 
equally among the three orthogonal directions: 
\begin{equation}
(\Delta v_{tan})^2={2\over 3}|\Delta {\bf v_{\rm rms}}(x)|^2,\quad {\rm and} \quad
(\Delta v_{rad})^2={1\over 3}|\Delta {\bf v_{\rm rms}}(x)|^2. 
\label{deltav_vec}
\end{equation}

The magnitude and direction of extra velocity of a typical dark matter particle 
in a given shell, eq.~\ref{deltav} and eq.~\ref{deltav_vec}, are the main 
quantitative results so far and will be used in the next 
section as the starting point for the halo collapse. As noted earlier,
the dark matter particles' positions are not updated by their respective 
displacements $d(x,t_c/2)$; we assume that these can be ignored because, being
randomly directed they average to zero for any small patch of the halo.
RG87 estimate that these displacements are small (see their Fig.9).

\subsection{Shell-crossing and equilibrium halo}\label{four}

Before we proceed with the collapse calculation we have to describe the core
of the halo, which was already introduced in Section~\ref{setup}. 
The density profile of the core can be chosen arbitrarily. 
The core is dynamically inert, i.e. once set its
density distribution is not changed throughout the collapse. The core is
required as the starting point, or seed, for the computation of the collapse 
of overlying shells. From the physical point of view one can imagine that the
core consists of the inner most shells that were the very first ones to 
collapse, and had enough time to come to an equilibrium before the 
computations described in this paper were started.

The collapse starts from the inner most shell, the one adjacent to the core. 
When the first shell reaches its $r_{max}$ it collapses and finds its 
$r_{apo}$ and $r_{peri}$ within the overall halo potential. It is assumed that
the potential is changing slowly compared to the dynamical timescales of the
shells, so that every shell conserves its adiabatic invariants, the radial and 
tangential momenta,
\begin{eqnarray}
j_\theta(x)=\Delta v_{tan}\; r_{max}\\
\label{jtheta}
j_r(x)=\int_{r_{peri}}^{r_{apo}} v_{rad}\, dr
\label{jr}
\end{eqnarray}
throughout the collapse. This is an important assumption in the RG87 formalism,
it is crucial to the computation of dynamics of shell crossing.

Up to the moment when a given shell reaches its maximum expansion radius
$r_{max}(x)$ at a time $t=t_c(x)/2$ corresponding to $\theta=\pi$, the shell
is assumed to be thin, its radial extent determined by the initial shell 
separation. At $r_{max}$, the average dark matter particle in the shell is 
given its additional random velocity, eq.~\ref{deltav} and 
eq.~\ref{deltav_vec}. Because the distribution of the secondary perturbations
in the halo is random the phase of $\Delta {\bf v_{\rm rms}}$ need not be the same for
all the particles in a given shell. In other words, at $r_{max}$ one particle
can get an outward kick (positive $\Delta v_{rad}$), while another, an
inward kick (negative $\Delta v_{rad}$) of the same rms amplitude. This makes
the shell `puff up' resulting in a radial distribution of dark matter particles of
a given shell between apocenter and pericenter, $r_{apo}$ and $r_{peri}$.
It is important to remember that the halos are assumed to be made of
collisionless dark matter particles with negligible
interaction cross-section, so that the particles can go through each 
other without generating shocks or dissipating energy and momentum. 

An alternative way of looking at the effect of $\Delta {\bf v_{\rm rms}}$ is 
that all particles of a given shell at $r_{max}$ got kicked `in phase'. 
Then the time average, taken over some reasonably long (but shorter than the
timescale of evolution of the shell) $\Delta t$, will show a distribution of 
particles in radii, again bounded by $r_{apo}$ and $r_{peri}$.

In either case the apocenter and pericenter can be calculated from the 
particle's energy integral and $\Delta v_{tan}$ and $\Delta v_{rad}$.
The energy integral,
\begin{equation}
E=\psi(r_{max})+{1\over 2}|\Delta {\bf v_{\rm rms}}|^2,
\label{energy}
\end{equation}
can be considered constant while a shell is still expanding. 

The radial velocity of a particle is
\begin{equation}
v_{rad}=[2(E-\psi(r))-(j_\theta/r)^2]^{1/2},\footnote{Note that RG87 define their
energy integral and potential as the negatives of the conventional definitions
of these quantities. We use the conventional definitions, i.e. potential is
negative quantity inside the halo, and energy integral of a bound particle
is negative.}
\label{vrad}
\end{equation}
Apocenter and pericenter are the radii where $v_{rad}$ becomes 0; at radii
outside of this range the radial velocity is imaginary. The radial distribution
of mass within a shell between apocenter and pericenter is not uniform. The
density in the radial range $dr$ around $r$ is proportional to the amount of 
time the particle spends there, (eq.[53] of RG87),
\begin{equation}
P(r)\;dr={{v_{rad}^{-1}\;dr}\over{\int_{r_{peri}}^{r_{apo}}v_{rad}^{-1}\;dr}}.
\label{probability}
\end{equation}

After the first shell has collapsed the halo has
a new radial density distribution, and so the halo's potential has 
to be recalculated before the second shell starts to collapse. The potential
contribution from the first shell is evaluated using eq.~\ref{probability}.
As the second shell collapses its pericenter will be closer to the center of 
the halo than the apocenter of the first shell, so the shells will overlap. 
This is the beginning of shell-crossing. As every additional shell falls in
and finds its apo- and pericenter (that satisfy its $j_\theta$, $j_r$ 
conservation)
the potential has to be recalculated, and hence the apo- and 
pericenters of all the interior shells that overlap the most recently fallen one.

As time goes by the gravitational potential interior to any given shell grows 
slightly deeper because of the collapse of the inner shells, and on average
every shell sinks further into the potential well. This goes on until the 
present time is reached.

\hoffset=-0.55in
{
\begin{deluxetable}{lllllll}

\tablewidth{0pc}
\tablecaption{Analytical Secondary Infall Models at a Glance}
\tablehead{
\colhead{Paper}  & Initial density & Cutoff  & Nonradial &  Self- & Conserved  & Final density\\
\colhead{}  & profile shape & radius & Motions & sim.? & quantities & profile 
}
\startdata
GG72 & $\delta$-fcn & none & none & yes & none & power law \\
G77  & $\delta$-fcn & none & none & yes & \underline {$J_r\!=\!r\, m(r)$} & power law\\
FG84 & \underline {power{\s}law} & none & none & yes & $J_r$ & power law \\
HS85 & \underline {corr.{\s}fcn.} & none & none & yes & $J_r$ & power law \\
RG87 & \underline {peaks{\s}} & none & \underline {sec. perturb.} & \underline{no{\s}} & $J_r$, \underline{$J_\theta$} & concave \\
ZH93 & power law & none & none & yes &$J_r$ & power law \\
AFH98& \underline{MAH{\s}} & none & \underline{thermal{\s}motions} & no & $J_r$, $J_\theta$ & concave\\
L00  & corr. fcn. & \underline {{\tiny {$1\over 2$}}$d_{\rm {peak-peak}}$} & none & no & $J_r$ & concave \\
N00  & power law & none & \underline{2 analyt. schemes} & yes & $J_r$, $J_\theta$ & concave \\

\tablerefs{
Gunn \& Gott (1972);
Gunn (1977);
Fillmore \& Goldreich (1984);
Hoffman \& Shaham (1985);
Ryden \& Gunn (1987);
Zaroubi \& Hoffman (1993);
Avila-Reese, Firmani \& Hernandez (1998);
Lokas (2000);
Nusser (2001).
}

\enddata
\label{table1}
\end{deluxetable}
}

\newpage

\end{document}